\documentclass[%
aps,
twocolumn,
superscriptaddress,
amsmath,amssymb,
prx,
floatfix
]{revtex4-2}

\usepackage{graphicx}
\usepackage{dcolumn}
\usepackage{bm}
\usepackage{enumitem}
\usepackage{xcolor}
\usepackage{physics}
\usepackage[hyperfootnotes=true]{hyperref} 

\newcommand{\rrangle}{\rangle\!\rangle}

\newcommand\identity{1\kern-0.25em\text{l}}

\makeatletter
\def\maketitle{
\@author@finish
\title@column\titleblock@produce
\suppressfloats[t]}
\makeatother

\begin{document}

\title{Universal Spreading of Conditional Mutual Information in Noisy Random Circuits}

\author{Su-un Lee}%
\email{suun@uchicago.edu}
\affiliation{%
 Pritzker School of Molecular Engineering, The University of Chicago, Chicago, IL 60637, USA
}%
\author{Changhun Oh}%
\affiliation{%
 Pritzker School of Molecular Engineering, The University of Chicago, Chicago, IL 60637, USA
}%
\affiliation{Department of Physics, Korea Advanced Institute of Science and Technology,
Daejeon 34141, Republic of Korea}
\author{Yat Wong}%
\affiliation{%
 Pritzker School of Molecular Engineering, The University of Chicago, Chicago, IL 60637, USA
}%
\author{Senrui Chen}%
\affiliation{%
 Pritzker School of Molecular Engineering, The University of Chicago, Chicago, IL 60637, USA
}%
\author{Liang Jiang}%
\email{liang.jiang@uchicago.edu}
\affiliation{%
 Pritzker School of Molecular Engineering, The University of Chicago, Chicago, IL 60637, USA
}%

\date{\today}

\begin{abstract} 
We study the evolution of conditional mutual information~(CMI) in generic open quantum systems, focusing on one-dimensional random circuits with interspersed local noise. Unlike in noiseless circuits, where CMI spreads linearly while being bounded by the lightcone, we find that noisy random circuits with an error rate $p$ exhibit superlinear propagation of CMI, which diverges far beyond the lightcone at a critical circuit depth $t_c \propto p^{-1}$. We demonstrate that the underlying mechanism for such rapid spreading is the combined effect of local noise and a scrambling unitary, which selectively removes short-range correlations while preserving long-range correlations. To analytically capture the dynamics of CMI in noisy random circuits, we introduce a coarse-graining method, and we validate our theoretical results through numerical simulations. Furthermore, we identify a universal scaling law governing the spreading of CMI.
\end{abstract}

\maketitle

\textit{Introduction}.---Understanding information spreading in non-equilibrium quantum systems is central to studying chaotic quantum systems, as well as essential for realizing quantum technologies. For these theoretical and practical purposes, random quantum circuits have served as a fruitful model to study the generic properties of quantum systems. The irreversible growth and spread of entanglement in random quantum circuits have elucidated the mechanisms underpinning quantum thermalization and chaos~\cite{deutsch_quantum_1991, srednicki_chaos_1994, nahum_quantum_2017, nahum_operator_2018}, provided useful models for black holes~\cite{hayden_black_2007, sekino_fast_2008, hosur_chaos_2016}, and showcased their utility in efficient quantum error-correcting codes~\cite{brown_short_2013, brown_decoupling_2015, gullans_quantum_2021}. However, incorporating noise is essential to enhance the physical relevance and applicability to near-term quantum devices. Thus, random quantum circuits interspersed with decoherence channels have been introduced to describe generic open quantum systems~\cite{noh_efficient_2020, li_entanglement_2023, zhang_information_2019, schuster_operator_2023}.

Among various measures characterizing information dynamics, conditional dependence holds fundamental and practical significance in quantum information and many-body physics.  It is intimately connected with the classical description of quantum states~\cite{molnar_approximating_2015, hastings_quantum_2007, leifer_quantum_2008, poulin_markov_2011}, efficient quantum algorithms for state preparations~\cite{brandao_finite_2019, kato_quantum_2019}, topological orders~\cite{kitaev_topological_2006,levin_detecting_2006, kim_conditional_2013, kim_universal_2023}, the information-theoretic measure of mixed-state quantum entanglement~\cite{christandl_squashed_2004, brandao_faithful_2011, seshadreesan_fidelity_2015}, and the quantum non-Markovianity~\cite{hayden_structure_2004, fawzi_quantum_2015, petz_monotonicity_2003, buscemi_information_2024}.

Given a state $\rho$, the direct correlation between the subsystems $A$ and $C$ is quantified by mutual information,
\begin{equation}
    I(A:C) = S(A)+ S(C) - S(AC),
\end{equation}
where $S(X) = -\Tr(\rho_X \log_2 {\rho_X})$ is the von Neumann entropy of the reduced density matrix $\rho_X$ of the subsystem $X$. Meanwhile, the dependence of $A$ and $C$, conditioned on another subsystem $B$, is captured by conditional mutual information~(CMI),
\begin{equation}
    I(A:C|B) = S(AB)+ S(BC) - S(B) - S(ABC).
\end{equation}
CMI is associated with approximate recoverability~\cite{fawzi_quantum_2015, petz_monotonicity_2003}, thereby serving as a measure of conditional dependence. The structures of CMI in the states in thermal equilibrium or topologically ordered states have been studied extensively~\cite{kato_quantum_2019, kuwahara_clustering_2020, harrow_classical_2020}. However, its behavior in dynamical systems remains less understood.

In this Letter, we study the dynamics of CMI in generic open quantum systems. Specifically, we consider quantum circuits that start with a product state and are subjected to random local gates, possibly with noise. We establish that in noiseless circuits, CMI spreads linearly constrained by the lightcone. Introducing noise, however, leads to the rapid spreading of CMI beyond the ligthcone. We demonstrate the underlying mechanism for such rapid spreading by showing that the local noise combined with a scrambling unitary selectively removes the short-range correlation while preserving the long-range correlation.

Equipped with the understanding of the non-local spreading of CMI, we study one-dimensional random circuits with interspersed local noise controlled by the error rate $p$. We introduce \textit{coarse-grained random circuit}, a new theoretical model that admits theoretical and numerical analysis. This model reveals that the higher the error rate within the circuit, the more rapid the spreading of CMI is. Remarkably, the spreading diverges at a critical timestep $t_c \propto p^{-1}$, leading to the entire system becoming conditionally dependent. Furthermore, we analytically identify and numerically validate a universal scaling law governing the spreading of CMI.

\begin{figure}
\includegraphics[width=\columnwidth]{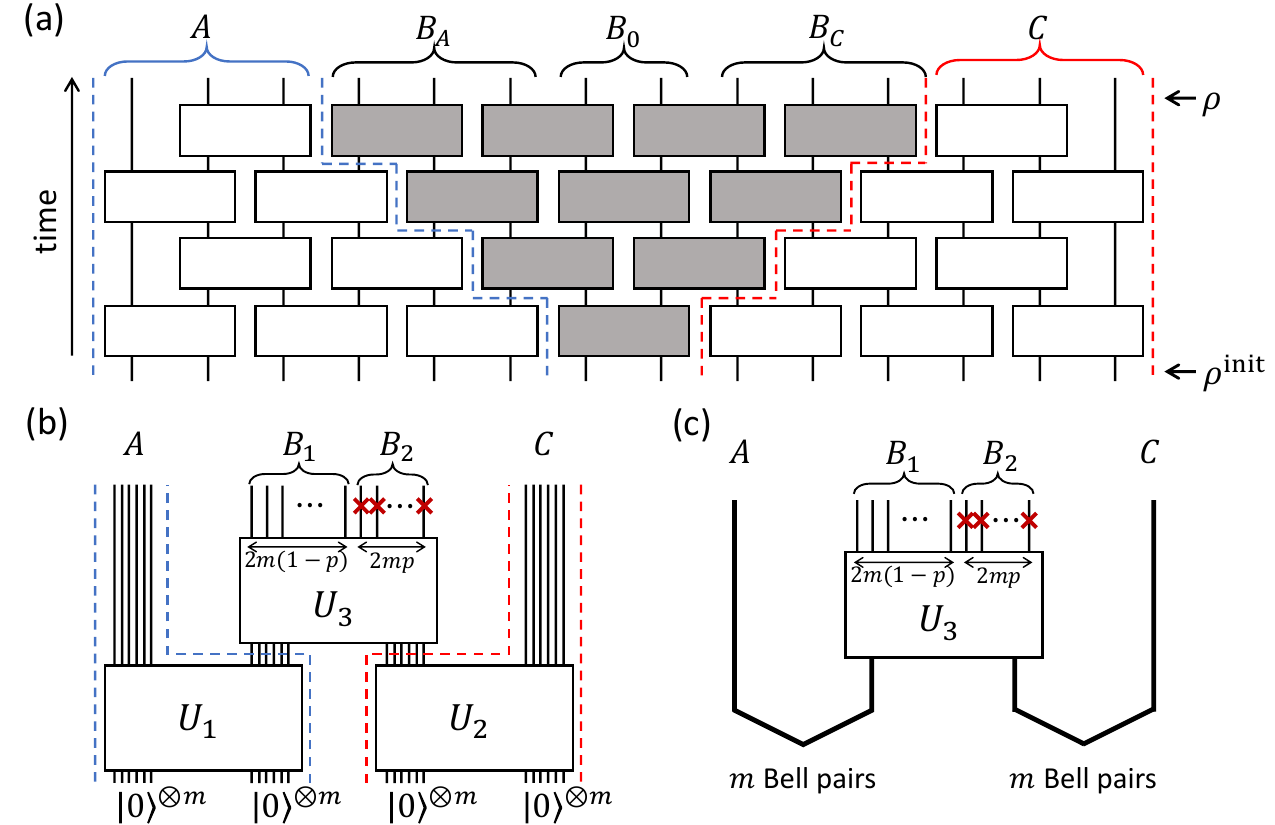}
\caption{\label{fig:PureAndMixed}(a) 1D noiseless quantum circuit. $\mathcal{L}(A)$ and $\mathcal{L}(C)$ are denoted by the regions enclosed by blue and red dashed lines, respectively. The gates outside of $\mathcal{L}(A) \cup \mathcal{L}(B)$ are colored grey. (b) Quantum circuit with noise. After two layers of random Clifford gates, a fraction of $p$ of the qubits in $B$ are completely depolarized. (c) Due to the gates $U_1$ and $U_2$ in (b), $A$ and $C$ effectively share $m$ Bell pairs with $B$, respectively.}
\centering
\end{figure}

\textit{Bound of lightcone in noiseless circuits.---}We first show that when a quantum circuit is noiseless, the spreading of CMI is bounded by the lightcone. Consider a one-dimensional quantum circuit of $N$ qubits that begins with a product state $\rho^{\mathrm{init}} = \bigotimes_{i = 1}^{N}\rho_i$. We choose two subsets of qubits $A$ and $C$, and denote $B$ to be ${\{1,2,\dots,N\}\setminus (A \cup C)}$ as depicted in Fig.~\ref{fig:PureAndMixed}(a). We say $A$ and $C$ are lightcone-separated if $\mathcal{L}(A) \cap \mathcal{L}(C)= \emptyset$, where $\mathcal{L}(X)$ is the backward lightcone of $X$.

Given the output state $\rho$, suppose $A$ and $C$ are lightcone-separated. We undo the gates outside of ${\mathcal{L}(A)\cup \mathcal{L}(C)}$ by applying the inverses of those gates successively. Since those are unitary gates applied only on $B$, they do not change $I(A:C|B)$. Then, the resultant state $\sigma$ after undoing those gates takes the form of ${\sigma = \sigma_{AB_A} \otimes \sigma_{B_0} \otimes \sigma_{CB_C}}$, where $B_{A}$ (or $B_{C}$) denotes the set of qubits that are in the support of $\mathcal{L}(A)$ (or $\mathcal{L}(C)$), and $B_0 = B \setminus (B_A \cup B_C)$.
This results in ${I(A:C|B)=0}$ from the definition of CMI. Therefore, the spreading of CMI is bounded by the lightcone in noiseless circuits.

\textit{Selective removal of correlation.---}Contrarily, noisy circuits can spread CMI into lightcone-separated regions. Here, we show that this rapid spreading stems from the fact that local noises, along with the scrambling unitary, selectively destroy the short-range correlation while preserving the long-range correlation. To understand this, we consider a quantum circuit of four blocks consisting of $m\gg 1$ qubits, as depicted in Fig.~\ref{fig:PureAndMixed}(b). We denote the first and last blocks by $A$ and $C$, respectively, and the two blocks in the middle by $B$. Starting from the product state $\ket{0}^{\otimes 4m}$, the two halves of the system undergo random Clifford gates $U_1$ and $U_2$, respectively. Then, we apply another random Clifford gate $U_3$ on $B$. Finally, we completely depolarize a fraction of $p$ of the qubits in $B$:
\begin{equation}
    \rho_{ABC} \mapsto \Tr_{B_2}\rho_{ABC} \otimes \frac{\identity_{B_2}}{d_{B_2}},
\end{equation}
where $B_2$ represents the depolarized qubits, and $B_1 = B\setminus B_2$. Here, $d_{B_2}$ is the Hilbert space dimension of $B_2$. Note that $A$ and $C$ are lightcone-separated.

A key observation is that depolarizing $B_2$ only removes the short-range correlation ${I(A:B)}$ but not the long-range correlation ${I(A:BC)}$, as long as ${p<1/2}$. Initially, random Clifford gates $U_1$ and $U_2$ almost maximally entangle the first and last two blocks because of the large dimensionality~\cite{preskill_lecture_2018, apel_holographic_2022}. This effectively forms $m$ Bell pairs shared by $A$ and $B$, and another $m$ Bell pairs shared by $B$ and $C$~[Fig.~\ref{fig:PureAndMixed}(c)]. Therefore, noting that each Bell pair counts two in mutual information, ${I(A:B)=I(B:C)=2m}$ and ${I(A:BC)=I(AB:C)=2m}$ before we depolarize $B_2$. 

Then, given $p<1/2$, the complete depolarization of $B_2$ preceded by $U_{3}$ decreases ${I(A:B)}$ by $4mp$. To see this, note that before the depolarization, $U_3$ nearly maximally entangle $AB_{1}$ with $B_{2}C$~\cite{preskill_lecture_2018, apel_holographic_2022}, resulting in $S(AB_{1}) = 2mp + m$. Meanwhile, ${S(A)=m}$ and ${S(B_1)=2m(1-p)}$ as both $A$ and $B_1$ are maximally entangled with the other subsystems. These leads to ${I(A:B_1)=2m(1-2p)}$, and thus,
\begin{equation}\label{eq:shortrange}
    I(A:B) = 2m(1-2p)
\end{equation}
after we depolarize $B_2$.

Contrarily, depolarizing $B_2$ does not decrease ${I(A:BC)}$. Before we depolarize $B_2$, the decoupling inequality~\cite{preskill_lecture_2018, abeyesinghe_mother_2009, hayden_decoupling_2008, hayden_black_2007} states that due to the scrambling unitary $U_3$, $A$ and $B_2$ are almost decoupled when ${p<1/2}$:
\begin{equation}\label{eq:decoupling}
    \mathbb{E}_{U_3} \left[ \left\lVert \rho_{AB_2}(U_3) - \rho_{A} \otimes \frac{\identity_{B_2}}{d_{B_2}} \right\rVert_1 \right] \le 2^{-m(1-2p)},
\end{equation}
where $\rho_{AB_2}(U_3)$ is the reduced density matrix of $AB_2$ after applying $U_3$. Here, $\lVert \cdot \rVert_1$ denotes the trace norm, and $\mathbb{E}_{U_3}$ represents averaging over random Clifford gates on $B$. The decoupling of $A$ and $B_2$ ensures the recovery of the $m$ Bell pairs with $A$ from $B_1 C$ with probability $1-\mathcal{O}(2^{-m(1-2p)})$~\cite{yoshida_efficient_2017, yoshida_recovery_2022}. Therefore, any operation on $B_2$ does not decrease ${I(A:BC)}$ in large $m$ limit. As a result, 
\begin{equation}\label{eq:longrange}
    I(A:BC) = 2m,
\end{equation}
after we depolarize $B_2$.

Eqs.~\eqref{eq:shortrange} and \eqref{eq:longrange} confirm that the loss of $B_2$ preceded by $U_3$ selectively destroys ${I(A:B)}$ while preserving ${I(A:BC)}$. Importantly, by the chain rule ${I(A:C|B)} = {I(A:BC)}-{I(A:B)}$, CMI is the discrepancy between the short-range and long-range correlations, resulting in ${I(A:C|B) = 4mp}$. Consequently, the lightcone-separated regions $A$ and $C$ share a non-zero CMI after the depolarization. See ~\footnote{See Supplemental Materials for the derivation of the analytical result, detailed numerical results, the protocol to generate Bell pairs between an arbitrary distance at $t \rightarrow t_c$, and the generalization to Haar random circuits with depolarizing channel, which includes Refs.~\cite{audenaert_entanglement_2005, li_measurement-driven_2019, sang_ultrafast_2023, cover_elements_2006, vidal_efficient_2003, verstraete_matrix_2004, zwolak_mixed-state_2004, bennett_mixed-state_1996, terhal_entanglement_2002, prosen_operator_2007, prosen_matrix_2009, xu_accessing_2020, kliesch_matrix_2014}}\label{suppl_mat} for an alternative derivation based on the stabilizer formalism.
\newcounter{supplmaterial}
\setcounter{supplmaterial}{\value{footnote}}

\begin{figure}
\includegraphics[width=\columnwidth]{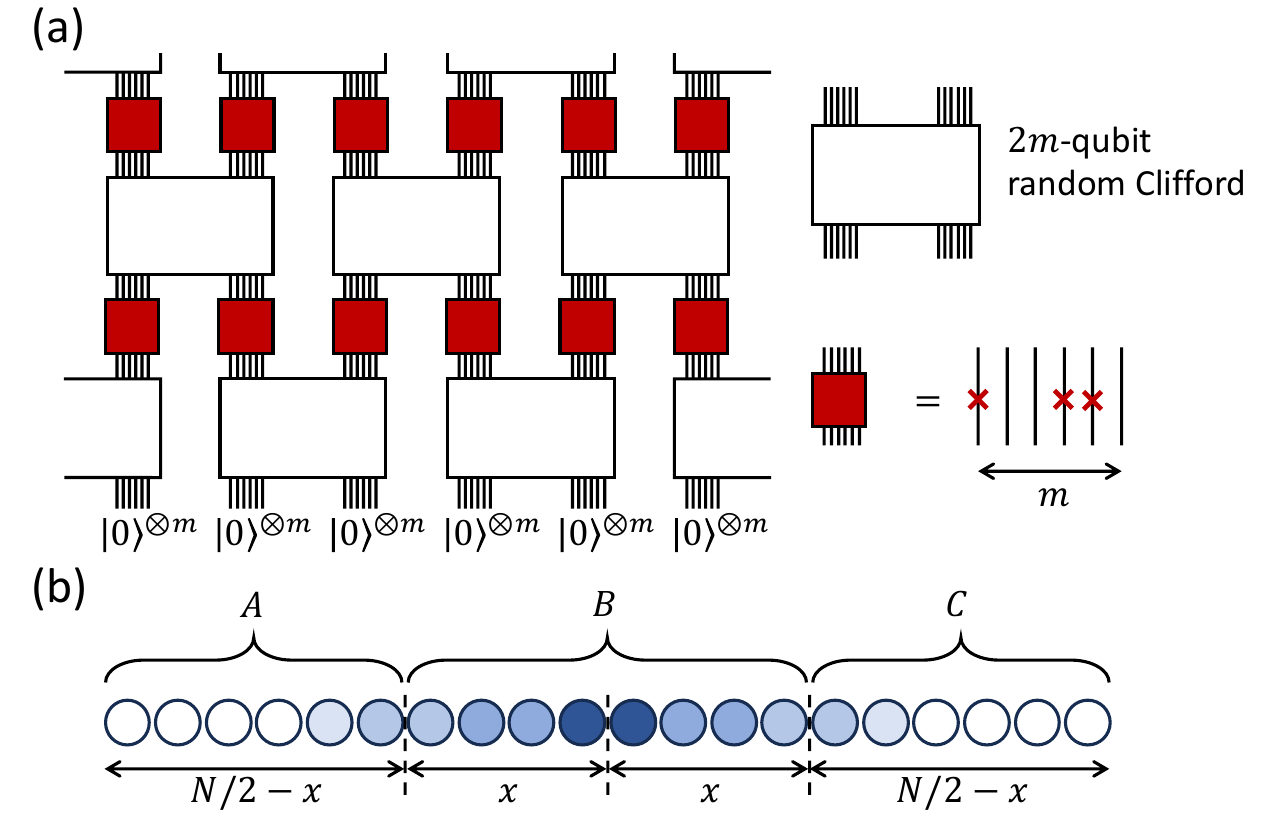}
\caption{\label{fig:Setup}(a) Coarse-grained random circuit. Beginning with a product state, we alternately apply random Clifford gates to the neighboring blocks, followed by heralded depolarizing channels. (b) At each timestep, the system is divided into $A$, $B$, and $C$ to calculate ${I^{\mathrm{norm}}(A:C|B)}$.}
\centering
\end{figure}

\begin{figure*}
\includegraphics[width=\textwidth]{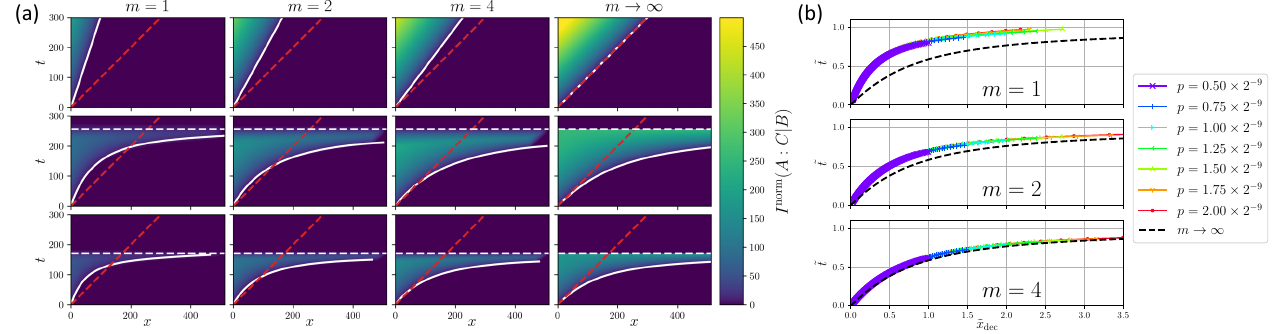}
\caption{\label{fig:numerical}(a) Evolution of ${I^{\mathrm{norm}}(A:C|B)}$ with various circuit depths $t$ and separation lengths $x$. The first three columns present the numerical results for the finite coarse-graining factors $m$, whereas the last column shows the analytical result from Eq.~\eqref{eq:I_beforerescalining}. The first row is the result of noiseless circuits ($p=0$), with subsequent rows for $p=1.0 \times 2^{-9}$ and $p=1.5 \times 2^{-9}$, respectively. Red dashed lines mark the limit set by lightcone ($x=t$). For noisy cases ($p>0$), $t_c$ is marked with white dashed lines. White solid lines represent $x_{\mathrm{dec}}(t)$. (b) $\tilde{x}_{\mathrm{dec}}(\tilde{t})$ with various error rates $p$ and coarse-graining factors $m$. The theoretical result for $m\rightarrow \infty$ is depicted as black dashed lines. The numerical results in (a) and (b) are based on averages from $1000$ circuit realizations with $N=2^{10}$.}
\centering
\end{figure*}

\textit{Coarse-grained random circuit}.---With the understanding of the non-local spreading of CMI, we study the dynamics of CMI in noisy random circuits. Inspired by Ref.~\cite{choi_quantum_2020}, we introduce the coarse-grained random circuit, a new model of noisy random circuits~[Fig.~\ref{fig:Setup}(a)]. This circuit comprises a one-dimensional array of $N \gg 1$ blocks, each containing $m$ qubits. Therefore, the coarse-graining factor $m$ determines the local dimensionality. Starting with a product state $\ket{0}^{\otimes mN}$, we alternately apply $2m$-qubit random Clifford gates on the even and odd pairs of the adjacent blocks. After each layer of unitary gates, we apply \textit{heralded depolarizing channel} on each qubit with an error rate $p$. A heralded depolarizing channel acting on qubit $i$ with an error rate $p$ is defined by completely depolarizing qubit $i$ with probability $p$:
\begin{equation}
\rho \mapsto
\begin{cases}
    \Tr_{i} \rho \otimes \identity_{i}/2 &\text{with Prob. of $p$,}\\
    \rho &\text{with Prob. of $1-p$,}
\end{cases}
\end{equation}
which is a simplified version of the depolarizing channel—this noise model is identical to the depolarizing channel, except the spacetime location of depolarization is determined for each circuit realization \footnotemark[\value{supplmaterial}]. Every timestep, we calculate the normalized CMI ${I^{\mathrm{norm}}(A:C|B)} = {I(A:C|B)/m}$, averaged over circuit realizations. Here $A$ is the first $N/2-x$ blocks, $B$ is the $2x$ blocks in the middle, and $C$ is the last $N/2-x$ blocks~[Fig.~\ref{fig:Setup}(b)]

The choice of unitary gates and noise channel is due to their classical simulability and analytical convenience, as Clifford gates and complete depolarization are efficiently described in the stabilizer formalism~\cite{gottesman_heisenberg_1998, aaronson_improved_2004}. Importantly, since the Clifford group forms a unitary 3-design~\cite{webb_clifford_2016, zhu_multiqubit_2017}, the averaged quantities are expected to be similar to those from the Haar random circuit or generic chaotic models~\cite{nahum_quantum_2017, nahum_operator_2018}. Moreover, we show in \footnotemark[\value{supplmaterial}] that the heralded depolarizing channel showcases qualitative similarity to the depolarizing channel when combined with random gates, supported by small-scale simulations of Haar random circuits with depolarizing channels.

Since the qubits are kept depolarized, the system's entropy increases until the state becomes maximally mixed. Roughly, each qubit is scrambled with $\sim m/p$ qubits before being depolarized. Thus, given a sufficiently small error rate, each qubit becomes maximally entangled with others before its depolarization. This results in two depolarized qubits for each depolarization event \footnote{If we depolarize one of the qubits of a Bell pair $\frac{1}{\sqrt{2}} (\ket{00}+\ket{11})$, both qubits are depolarized}. Therefore, at timestep $t$, $~m(1-2pt)$ qubits remain un-depolarized in each block, and the system becomes almost maximally mixed at the critical timestep $t_c = 1/2p$.

Fig.~\ref{fig:numerical}(a) presents the evolution of ${I^{\mathrm{norm}}(A:C|B)}$ for finite $m$ (numerical) and $m \rightarrow \infty$ (analytical) with various error rates. In noiseless circuits, ${I(A:C|B)=0}$ for $x>t$, since $A$ and $C$ are lightcone-separated. We observe that the spreading of $I(A:C|B)$ becomes faster as we increase the coarse-graining factor $m$. This is because there are non-negligible portions of Clifford gates that do not maximally entangle the two blocks when $m$ is small, and that portion shrinks as $m$ grows~\cite{preskill_lecture_2018, apel_holographic_2022}.

Contrarily, introducing noise results in ${I(A:C|B)}$ attaining non-zero values in regions well beyond the lightcone, attributed to the selective removal of short-range correlations. This phenomenon can be analytically captured in the limit $m \rightarrow \infty$~\footnotemark[\value{supplmaterial}]. As seen in Eq.~\eqref{eq:longrange}, depolarization channels cannot decrease the long-range correlation ${I(A:BC)}$ as long as less than half of the qubits are depolarized. Meanwhile, at every two timesteps, a $2m$-qubit random Clifford gate is applied between $A$ and $B$, thereby increasing ${I(A:BC)}$. Since the random Clifford gate maximally generates entanglement in the large $m$ limit and that there are $2m(1-2pt)$ non-depolarized qubits on its support, ${I(A:BC)}$ increases by ${4m(1-2pt)}$ at every two timesteps. Therefore, by integrating this increment,
\begin{equation}
    I(A:BC) = 2mt(1-pt),
\end{equation}
for $t<t_c$. In contrast, using a similar method for deriving Eq.~\eqref{eq:shortrange}, we show in \footnotemark[\value{supplmaterial}] that the short-range correlation ${I(A:B)}$ is upper bounded by ${2m(1-2pt)x}$, resulting in
\begin{equation}\label{eq:I_AB}
    I(A:B) = \min\{2m(1-2pt)x, I(A:BC)\}.
\end{equation}
Therefore, by the chain rule,
\begin{equation}\label{eq:I_beforerescalining}
    I^{\mathrm{norm}}(A:C|B) = \max\{2t(1-pt)-2(1-2pt)x,0\},
\end{equation}
given $t<t_c$. Note that $I(A:C|B)=0$ when $t > t_c$ as the system becomes maximally mixed.

As we increase the separation $x$, ${I^{\mathrm{norm}}(A:C|B)}$ decays until it becomes zero. We define $x_{\mathrm{dec}}(t)$ as the separation $x$ at which ${I(A:C|B)}$ begins to be zero for $t<t_c$, indicating the spreading front of CMI. Then, Eq.~\eqref{eq:I_beforerescalining} demonstrates that
\begin{equation}\label{eq:xdec_beforerescalining}
    x_{\mathrm{dec}}(t) = \frac{t(1-pt)}{1-2pt}.
\end{equation}
One can see that the spreading becomes more rapid as we increase the error rate $p$. Notably, the spreading front $x_{\mathrm{dec}}(t)$ diverges as ${t \rightarrow t_c}$. Therefore, a non-negligible amount of conditional dependence spreads out to the entire system right before the state becomes maximally mixed.

Our derivations of the rapid spreading and critical behavior of CMI are based on the limit of $m\rightarrow \infty$, where the long-range correlation is perfectly protected from the local noise while the short-range correlation keeps decreasing. However, we emphasize that this selective removal of short-range correlation also emerges in the cases of finite $m$ as long as the error rate is sufficiently low. As seen in Fig.~\ref{fig:numerical}(a), those features of rapid spreading and divergence at $t_c$ are observed in the numerical results of the finite $m$.

\textit{Universal scaling law}.---Remarkably, the dynamics of CMI exhibit a universal scaling law. Specifically, we consider the following rescaling:
\begin{align}
    \tilde{t} &= 2pt, \label{eq:rescaling_t}\\
    \tilde{x} &= 2px, \label{eq:rescaling_x}\\
    \tilde{I}^{\mathrm{norm}}(A:C|B) &= pI^{\mathrm{norm}}(A:C|B). \label{eq:rescaling_I}
\end{align}
This rescaling makes Eqs.~\eqref{eq:I_beforerescalining} and \eqref{eq:xdec_beforerescalining} error-rate-independent, demonstrating the universal scaling law:
\begin{equation}\label{eq:I_afterrescalining}
    \tilde{I}^{\mathrm{norm}}(A:C|B) = \max\{0, \tilde{t}(1-\tilde{t}/2) - (1-\tilde{t})\tilde{x}\},
\end{equation}
\begin{equation}\label{eq:xdec_afterrerescalining}
    \tilde{x}_{\mathrm{dec}}(\tilde{t}) = \frac{\tilde{t}(1-\tilde{t}/2)}{1-\tilde{t}},
\end{equation}
for $\tilde{t}<1$, where $\tilde{x}_{\mathrm{dec}}(\tilde{t}) = 2px_{\mathrm{dec}}(\tilde{t}/2p)$. Fig.~\ref{fig:numerical}(b) presents rescaled spreading fronts $\tilde{x}_{\mathrm{dec}}(\tilde{t})$ from the numerical results of finite $m$ along with the analytical result of $m\rightarrow \infty$ in Eq.~\eqref{eq:xdec_afterrerescalining}. For a fixed coarse-graining factor $m$, the spreading fronts across varying error rates collapse into a single curve, affirming the applicability of the universal scaling law even for the circuit with finite $m$. As we increase $m$, the collapsed curve $\tilde{x}_{\mathrm{dec}}(\tilde{t})$ approaches the analytical result. 

Note that slight deviations exist in the rescaled spreading fronts $\tilde{x}_{\mathrm{dec}}(\tilde{t})$ for large $p$ values when $m=1$. In the circuits with finite $m$, local noise still selectively removes short-range correlations ${I(A:B)}$. However, it can reduce ${I(A:BC)}$ for a small amount since the correlation is scrambled into a finite number of qubits. This introduces higher-order effects of $p$ on ${I(A:BC)}$, leading to small deviations in $\tilde{x}_{\mathrm{dec}}(\tilde{t})$. Fig.~\ref{fig:numerical}(b) shows that such higher-order effects are suppressed as we decrease the $p$ or increase $m$.

\textit{Discussion}.---Rapid spreading of CMI shows that causally separated regions become conditionally dependent. However, it is important to remark that although the non-trivial correlation extends far beyond the lightcone, our findings do not conflict with the Lieb-Robinson bound or the principles of causality~\cite{lieb_finite_1972, bravyi_lieb-robinson_2006, chen_speed_2023}, as local observers in $A$ and $C$ cannot detect $I(A:C|B)$.

Meanwhile, the divergence of $x_{\mathrm{dec}}(t)$ as $t \rightarrow t_c$ indicates that CMI spreads into the entire system. We emphasize that this pervasive conditional dependence is not sorely composed of classical correlations but the multipartite quantum correlation extending throughout the entire system. Specifically, at the critical timestep, we show in ~\footnotemark[\value{supplmaterial}] that no matter how largely separated $A$ and $C$ are, measuring $B$ and heralding the measurement outcome to $A$ and $C$ generate Bell pairs shared by $A$ and $C$. Here, CMI acts as an upper bound of the number of generatable Bell pairs.

This result bears similarities to entanglement swapping~\cite{bennett_teleporting_1993}. In the entanglement-swapping protocol, two separate Bell pairs between $A$, $B$, and $B$, $C$ are prepared, and the measurement on $B$ generates entanglement between $A$ and $C$. In our case, however, we have inseparable multipartite correlations among $A$, $B$, and $C$, and the measurement on $B$ reduces them into entanglement between $A$ and $C$~\footnotemark[\value{supplmaterial}].

Finally, we remark that prior studies have shown that there exist constant depth circuits acting on a product state that generate CMI between arbitrarily large distant regions~\cite{zou_spurious_2016, williamson_spurious_2019, kato_toy_2020, kim_universal_2023}. Our findings further extend this understanding by demonstrating that the formation of long-distance conditional dependence also emerges within noisy random circuits without designing a sophisticated circuit to achieve it.

\textit{Note added}.---Upon completion of our study, we became aware of related independent work~\cite{zhang_nonlocal_2024}, which we believe complements our findings.

We acknowledge support from the ARO(W911NF-23-1-0077), ARO MURI (W911NF-21-1-0325), AFOSR MURI (FA9550-19-1-0399, FA9550-21-1-0209, FA9550-23-1-0338), DARPA (HR0011-24-9-0359, HR0011-24-9-0361), NSF (OMA-1936118, ERC-1941583, OMA-2137642, OSI-2326767, CCF-2312755), NTT Research, Packard Foundation (2020-71479), and the Marshall and Arlene Bennett Family Research Program. This material is based upon work supported by the U.S. Department of Energy, Office of Science, National Quantum Information Science Research Centers. S.L. is partially supported by Kwanjeong Educational Foundation. The authors are also grateful for the support of the University of Chicago’s Research Computing Center for assistance with the numerical experiments carried out in this work.

\bibliographystyle{apsrev4-2}
\bibliography{references_mod}

\clearpage

\setcounter{equation}{0}
\setcounter{figure}{0}
\setcounter{table}{0}
\setcounter{page}{1}
\setcounter{section}{0}
\makeatletter
\renewcommand{\theequation}{S\arabic{equation}}
\renewcommand{\thefigure}{S\arabic{figure}}
\renewcommand{\bibnumfmt}[1]{[S#1]}

\title{Supplemental Material for ``Universal Spreading of Conditional Mutual Information in Noisy Random Circuits"}
\maketitle

\renewcommand{\theequation}{S\arabic{equation}}
\renewcommand{\thefigure}{S\arabic{figure}}

\onecolumngrid

\section{\label{sec:stabilizer}Preliminaries}

\subsection{\label{subsec:stabilizer}Stabilizer formalism}

Stabilizer formalism provides a compact and efficient way to represent a wide class of quantum states and their dynamics, utilizing a stabilizer group \cite{gottesman_heisenberg_1998, aaronson_improved_2004}. A stabilizer group $\mathcal{S}$ is an Abelian subgroup of the Pauli group $\mathcal{P}_N$ of $N$ qubits with the constraint of $-I \notin \mathcal{S}$. Let $G = \{ g_1, g_2, \cdots, g_K \}$ be a generating set of the stabilizer group $\mathcal{S}$, i.e., $\mathcal{S} = \langle g_1, g_2, \cdots, g_K \rangle$, where $K$ is the number of the generators. A stabilizer state $\rho$ corresponding to $\mathcal{S}$ is a uniform mixture over the vector space spanned by $2^{N-K}$ simultaneous eigenstates of $\mathcal{S}$. The state $\rho$ is explicitly written as
\begin{equation}\label{eq:explicit_stab}
    \rho = \frac{1}{2^N}\sum_{g \in \mathcal{S}} g = \frac{1}{2^{N-K}} \prod_{k=1}^{K} \frac{\identity+g_k}{2}.
\end{equation}
Therefore, we can efficiently express the dynamics of the state $\rho$ only by keeping track of the generating set $G$. Since $\rho$ is a uniform mixture over $2^{N-K}$ orthonormal states, the von Neumann entropy $S_{\rho} = -\Tr(\rho\log_2\rho)$ is $N-K$. Note that $\rho$ is a pure state when $K = N$ and a maximally mixed state when $K = 0$.

For notational convenience, we introduce a \textit{stabilizer tableau} to represent $G$, which is a matrix of $K$ rows and $N$ columns. The $k$-th row is assigned to the generator $g_k \in G$, and in that row, the element at $i$-th column is the Pauli operator applied by $g_k$ to the $i$-th qubit. To have a complete description of $G$, an additional array of $K$ entities is needed to store the overall phases of each generator. However, we disregard these phases as they are irrelevant to the amount of correlation in the state. Additionally, for a stabilizer group $\mathcal{S}$, note that the corresponding stabilizer tableau is not unique, introducing a gauge degree of freedom. One such example is row permutation; the stabilizer tableau still represents the same generating set $G$ after the rows are permuted. Another degree of freedom is row multiplication, where multiplying one row by another alters the generating set, but the resultant generating set still generates the same stabilizer group $\mathcal{S}$.

When a Clifford gate $U$ is applied to the state $\rho$, we simply update each row from $g_k$ to $U g_k U^{\dagger}$, since $G' = \{U g_1 U^{\dagger}, U g_2 U^{\dagger}, \cdots, U g_K U^{\dagger}\}$ represents the resultant state $U \rho U^{\dagger}$. Another useful operation is complete depolarization. Complete depolarization of the qubit $i$ maps the state $\rho$ into $\Tr_{i}\rho \otimes \identity_{i}/2$, where $\identity_{i}$ is the identity operator acting on the qubit $i$. To apply complete depolarization to the qubit $i$ for a given stabilizer tableau, there are three cases to consider:
\begin{enumerate}[leftmargin=*,align=left]
    \item[\textit{Case 1.}] Every element of the $i$-th column is $\identity$. In this case, the complete depolarization does not change the stabilizer tableau.
    \item[\textit{Case 2.}] If there is only one type of nontrivial Pauli operator on the $i$-th column, suppose $k$-th row contains this nontrivial Pauli operator at the $i$-th column. Then, perform row reduction by multiplying other rows with the $k$-th row, ensuring every row except for the $k$-th row has $\identity$ on the $i$-th column. Finally, remove the $k$-th row.
    \item[\textit{Case 3.}] If there are two or more kinds of nontrivial Pauli operators on the $i$-th column, let the $k_1$-th and $k_2$-th row have different nontrivial Pauli operators at the $i$-th column.  Perform the row reduction by multiplying other rows with the $k_1$-th and $k_2$-th row, ensuring every row except those two rows has $\identity$ on the $i$-th column. Finally, remove the $k_1$-th and $k_2$-th rows.
\end{enumerate}
In summary, complete depolarization reduces the number of rows, and the extent of this reduction varies, thereby increasing the entropy. See Ref. \cite{audenaert_entanglement_2005} for a detailed explanation of complete depolarization in the stabilizer formalism.

The last useful operation introduced here is a measurement with a Pauli observable. Suppose a generating set $G = \{g_1, g_2, \dots, g_K\}$ of a stabilizer group $\mathcal{S}$ that corresponds to a stabilizer state is given. After the measurement with a Pauli observable $g$, we update $G$ as follows:
\begin{enumerate}[leftmargin=*,align=left]
    \item[\textit{Case 1.}] If $g \in \mathcal{S}$ or $-g \in \mathcal{S}$, then $G$ is unaffected. The measurement outcome is $+1$ when $g \in \mathcal{S}$, and otherwise, the outcome is $-1$.
    \item[\textit{Case 2.}] If $[g,g_i] = 0$ for all $g_i \in G$ but $g \notin \mathcal{S}$, then the measurement outcome is $+1$ or $-1$ with equal probability. If the outcome is $\pm1$, then we add $\pm g$ in $G$: $G = \{\pm g, g_1, g_2, \dots, g_K\}$.
    \item[\textit{Case 3.}] If $g$ anticommutes with some of the generators in $G$, then the measurement outcome is $+1$ or $-1$ with equal probability. Let $g_1$ be one of such generators. By multiplying the other generators that are anticommuting with $g$ by $g_1$, we make $g_1$ the only generator that anticommutes with $g$. Then, if the outcome is $\pm 1$, we replace $g_1$ in $G$ with $\pm g$: $G=\{\pm g, g_2, \dots, g_K\}$.
\end{enumerate}
Note that for the last two cases, the measurement outcome only determines the overall phase of the generator, which is disregarded in this work.

\subsection{Clipped gauge}

As mentioned, two different generating sets can generate the same stabilizer group, and thus, there is a gauge degree of freedom in choosing the corresponding generating set $G$. The \textit{clipped gauge} is a specific choice of generating set that is useful for analyzing correlations in one-dimensional systems. The clipped gauge is first introduced in Ref. \cite{nahum_quantum_2017}, with detailed explanations available in the Appendix of Ref. \cite{li_measurement-driven_2019}.

Consider an $N$-qubit system arranged in a one-dimensional array, and suppose a generating set $G$ that corresponds to an $K \times N$ stabilizer tableau is given. For $k=1,2,\dots,K$, define the left (or right) endpoint $l(k)$ (or $r(k)$) as the index of the first (or last) column that has a non-trivial element on the $k$-th row. Additionally, define $\rho_l(i)$ (or $\rho_r(i)$) as the number of rows that have $i$ as their left (or right) endpoint. The choice of the generating set $G$ of the stabilizer group is said to be in the clipped gauge if, for $i = 1, 2, \cdots, N$,
\begin{enumerate}
    \item[(i)] $\rho_l(i) + \rho_r(i) \le 2$, and
    \item[(ii)] if $\rho_{l}(i)=2$ or $\rho_r(i)=2$, then the two rows that have their endpoint at $i$ should have different Pauli operators on the $i$-th column.
\end{enumerate}
Importantly, it is possible to achieve these two conditions of the clipped gauge for any stabilizer state. The clipping algorithm consists of two parts: First, we make the stabilizer tableau in the row-reduced echelon form (RREF) through Gaussian elimination \cite{audenaert_entanglement_2005}. Secondly, we transform the tableau into a lower triangular form by performing row-reduction on the right endpoints without affecting the left endpoints. More explicitly, starting from $i=K$ up to $i=2$, we utilize the $i$-th row to eliminate the $r(i)$-th column of $j$-th rows for $j<i$ \cite{li_measurement-driven_2019}. For analytical convenience, we permute the rows so that $l(i)+r(i) \le l(j) + r(j)$ for $i < j$ so that the mid-point is monotonically increasing with the row index.

The clipped gauge is particularly useful when studying correlations in stabilizer states. Specifically, for two neighboring contiguous regions $A$ and $B$, with $A$ on the left and $B$ on the right, it is known that \cite{sang_ultrafast_2023}
\begin{equation}\label{eq:Clipped_MI}
    I(A:B) = |\{k : l(k) \in A \text{ and } r(k) \in B\}|.
\end{equation}
Furthermore, consider two contiguous regions $A$ and $C$ that are separated by another region $B$. Then, by applying the chain rule of mutual information $I(A:BC) = I(A:B)+I(A:C|B)$, we have
\begin{equation}\label{eq:Clipped_CMI}
    I(A:B|C) = |\{k : l(k) \in A \text{ and } r(k) \in C\}|.
\end{equation}
Given a stabilizer tableau in the clipped gauge, we define the length of the $k$-th stabilizer generator as
\begin{equation}
    len(k) = r(k) - l(k) + 1.
\end{equation}
Then, by Eq.~\eqref{eq:Clipped_CMI}, we see that a non-zero conditional mutual information (CMI) between largely separated regions means the presence of stabilizer generators with a large length in the clipped gauge.

\section{\label{sec:coarse-graining}Derivation of the analytical result}

The analytical results presented in the main text utilize the coarse-graining method, applied within the limit of a large coarse-graining factor. In this section, we explain how the coarse-graining method is applied to derive these analytical results under a plausible assumption. Initially, we introduce an assumption regarding the stabilizer tableau of a random stabilizer state, termed the \textit{maximal scrambling ansatz}, and subsequently numerically verify this assumption. Employing the coarse-graining method in conjunction with the maximal scrambling ansatz significantly simplifies the dynamics of the tableau, thereby facilitating analytical calculations. Using these methods, we derive the results for the four-block example and the coarse-grained random circuit, as presented in the main text.

\subsection{Random stabilizer state and maximal scrambling ansatz}

\begin{figure*}
\includegraphics[width=\textwidth]{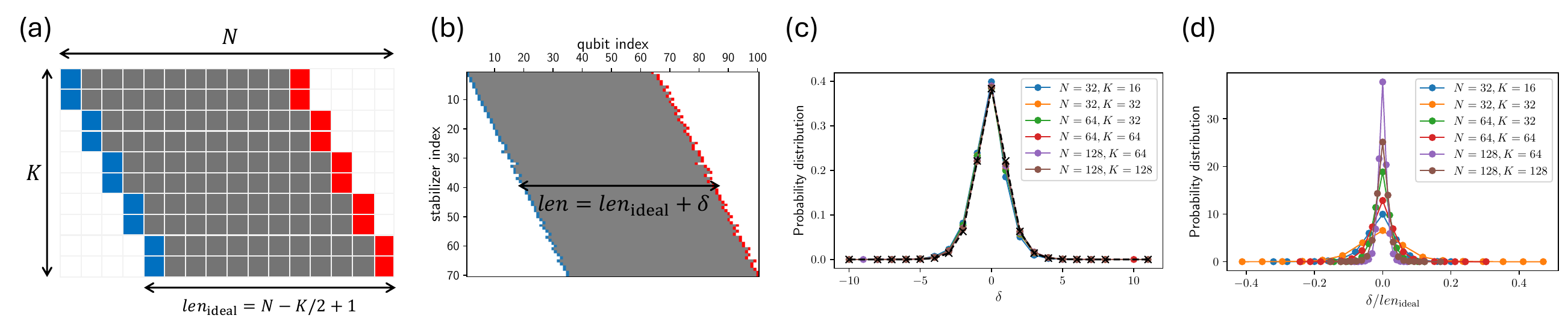}
\caption{\label{fig:random_stabilizer} (a) Stabilizer tableau of the maximal scrambling ansatz. Each row has left and right endpoints marked in blue and red, respectively. The elements between these endpoints are grey, and the elements outside the endpoints are $\identity$ and are colored white. Each row has the length of $len_{\mathrm{ideal}}=N-K/2+1$. (b) An example of a clipped gauge of a random stabilizer state of $N=100$ qubits with $K=70$ stabilizer generators. The length of each row is close to that of maximal scrambling ansatz $len_{\mathrm{idal}}$ with a small fluctuation $\delta$. (c) Distribution of the lengths of stabilizers of random stabilizer states in clipped gauge. The distributions of various values of $N$ and $K$ are presented, and each distribution is averaged over $1000$ random stabilizer states. The deviation of the length of the stabilizer generators from that of the maximal scrambling ansatz $len_{\mathrm{ideal}}$ is regardless of the values $N$ and $K$. The black dashed line denotes the fitted line $y=A\sech^2(Bx)$ from the data with $N=128$ and $K=128$, indicating that a large deviation is exponentially unlikely. (d) Distribution of $\delta/len_{\mathrm{ideal}}$. Since the deviation of the length of the stabilizer generators from $len_{\mathrm{ideal}}$ is independent of $N$ and $K$, it is negligible when compared to the value of $len_{\mathrm{ideal}}$ in the large $N$ limit.}
\centering
\end{figure*}

When we assume a large coarse-graining factor and apply random Clifford gates, we encounter large stabilizer tableaus that consist of random stabilizers. Here, we make a plausible assumption about the stabilizer tableau of random stabilizers and numerically verify it. Consider a random stabilizer state of $N$ qubits, which consists of $K$ stabilizer generators $g_1, g_2, ..., g_K$ ($K\le N$) that are randomly chosen assuming $N\gg1$ and $K\gg1$. Consequently, the rows of the stabilizer tableau of this state are random Pauli strings such that (1) $g_1, g_2, ..., g_K$ are independent, and (2) $[g_i,g_j]=0$ for any $i,j = 1,2, ..., K$. Note that this random stabilizer state is equivalent to a stabilizer state after the application of an $N$-qubit random Clifford gate on any stabilizer state of $N$ qubits and $K$ stabilizer generators.

Random Clifford gates on a large number of qubits maximally generate entanglement between any bipartition \cite{preskill_lecture_2018, apel_holographic_2022}. This motivates us to assume that the clipped gauge of the random stabilizer state is in the \textit{maximal scrambling ansatz}. The maximal scrambling ansatz for $N$ qubits and $K$ stabilizer generators denotes a stabilizer tableau in the clipped gauge comprising stabilizer generators whose lengths are all $len_{\mathrm{ideal}}=N+K/2+1$. Given that every column is required to have no more than two endpoints in the clipped gauge, the stabilizer tableau of the maximal scrambling ansatz has a parallelepiped shape, as depicted in Fig.~\ref{fig:random_stabilizer}(a). Notably, for $i \le N/2$, the bipartitions $A = \{1,2, \dots,i\}$ and $B = \{i+1, i+2, \dots, N\}$ exhibit the maximal mutual information of
\begin{equation}
I(A:B) = \min\{2i,K\},
\end{equation}
by Eq.~\eqref{eq:Clipped_MI}. If we select $A$ and $B$ to be non-contiguous regions, they can be rendered contiguous by permuting the qubits, and the random stabilizer state remains invariant under such permutation. Therefore, the maximal scrambling ansatz demonstrates the maximal correlation between any bipartition.

Fig.~\ref{fig:random_stabilizer}(b) presents an example of a clipped gauge from a random stabilizer state with $N=100$ qubits and $K=70$ stabilizer generators. Here, the lengths of the stabilizer generators approximate those predicted by the maximal scrambling ansatz, with a minor deviation $\delta$:
\begin{equation}
    len = len_{\mathrm{ideal}} + \delta.
\end{equation}
Specifically, Fig.~\ref{fig:random_stabilizer}(c) illustrates the distribution of the lengths of the stabilizer generators. As the distributions for various values of $N$ and $K$ closely align, this suggests that the deviation $\delta$ is independent of $N$ and $K$. Furthermore, the distribution of $\delta$ matches well with the fitting line $y = A\sech^2(Bx)$, where $A = 0.3835$ and $B = 0.7762$. This indicates that a large deviation is exponentially unlikely. Therefore, the deviation $\delta$ becomes negligible compared to $len_{\mathrm{ideal}}$ as $N$ increases, as demonstrated in Fig.~\ref{fig:random_stabilizer}(d). In the derivation of the analytical results in the main text, such fluctuations negligible to the size of the stabilizer do not affect the results. Consequently, we adopt the maximal scrambling ansatz for random stabilizer states in subsequent sections.

\subsection{Four-block example}

\begin{figure}
\includegraphics[width=\textwidth]{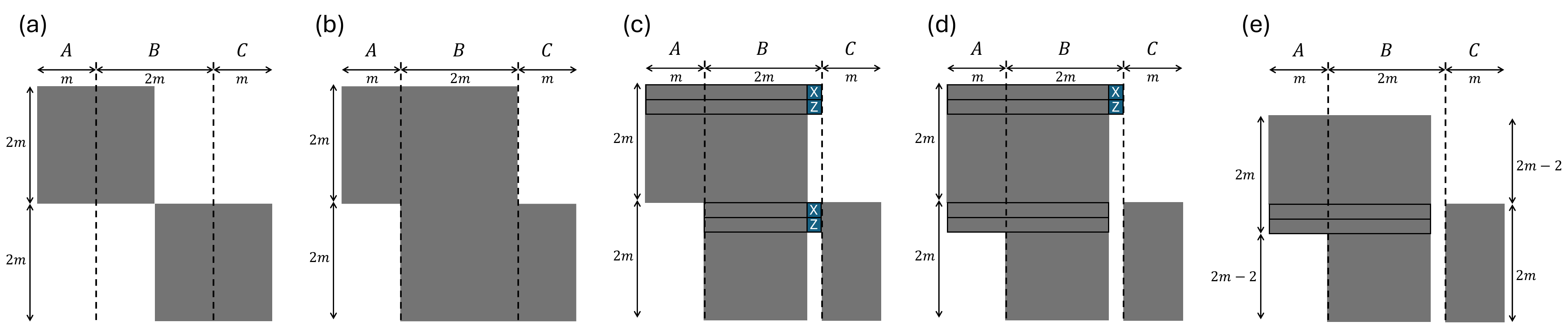}
\caption{\label{fig:4blocks}
(a) Stabilizer tableau after applying $U_1$ and $U_2$. The large random Clifford gates $U_1$ and $U_2$ make the two $2m \times 2m$ block diagonals consist of random stabilizers. The random tableau entities are colored grey, and the trivial entities of $\identity$'s are colored white. (b) Stabilizer tableau after applying $U_3$. $U_3$ makes the rows that act nontrivially on $B$ extend to the entire $B$. (c) We find two independent Pauli operators on the $3m$-th column among the first $2m$ rows and bring them to the first two rows. Then, by doing row reductions using those rows, we make the other elements of the $3m$-th column all $\identity$'s for the first $2m$-rows. We do the same task on the last $2m$ rows. (d) Stabilizer tableau after multiplying the $(2m+1)$-th and $(2m+2)$-th rows by the first and second rows, respectively. (c) Removing the first two rows completes the complete depolarization on $3m$-th qubit. This results in the loss of two rows while preserving the number of rows that randomly act on $A$ and those randomly act on $C$.}
\centering
\end{figure}

\begin{figure}
\includegraphics[width=\textwidth]{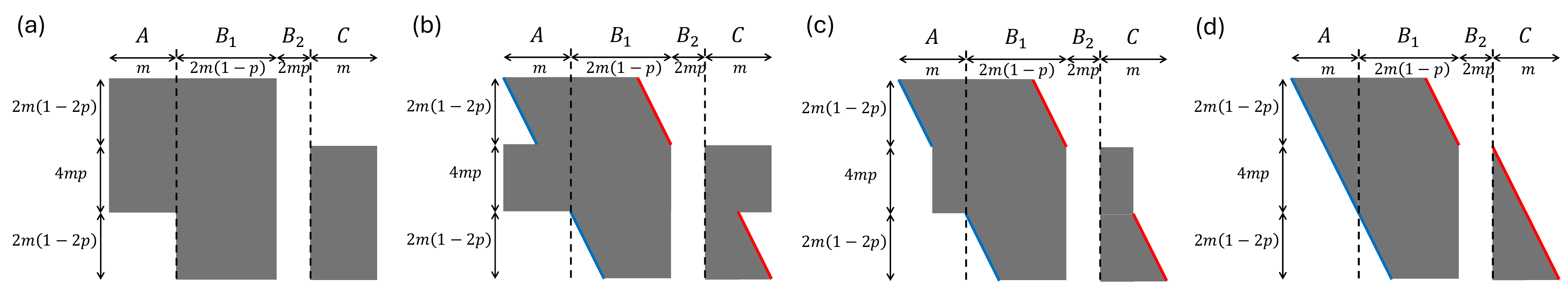}
\caption{\label{fig:4blocks_clipped}Procedure of taking clipped gauge after completely depolarizing the qubits in $B_2$. (a) Stabilizer tableau after depolarizing all qubits in $B_2$. (b) We take maximal scrambling ansatz for the submatrices of the first $2m(1-2p)$ rows and the last $2m(1-2p)$ rows. Here, the left and right endpoints are colored blue and red, respectively. Since we take the limit of $m \gg 1$, the maximal scrambling ansatz approaches the parallelepiped shape. (c) By row reduction with the first and last $2m(1-2p)$ rows, we make the first and last $m(1-2p)$ columns of the $4mp$ rows in the middle all $\identity$'s (d) Finally, we take maximal scrambling ansatz on the submatrix of the $4mp$ rows in the middle. This results in the stabilizer tableau in the clipped gauge.}
\centering
\end{figure}

Equipped with the assumption of the maximal scrambling ansatz, we derive the results of the four-block example described in Fig.~2(a) in the main text. The approach of the maximal scrambling ansatz reproduces Eqs.~(4) and (6) in the main text. In the four-block example, we consider a quantum circuit of four blocks of qubits, each consisting of $m$ qubits with $m \gg 1$. We denote the first and last blocks as $A$ and $C$, respectively, and the two in the middle as $B$. The circuit starts from the product state $\ket{0}^{\otimes 4m}$, corresponding to the $4m \times 4m$ stabilizer tableau of all $Z$’s on its diagonal and $\identity$’s elsewhere. Then, we apply a $2m$-qubit random Clifford gate $U_1$ on $A$ and the left half of $B$, and another $2m$-qubit random Clifford gate $U_2$ on the other half of $B$ and $C$. The random Clifford gates $U_1$ and $U_2$ fill the two $2m \times 2m$ block diagonals with random stabilizers, as seen in Fig.~\ref{fig:4blocks}(a). After that, we apply another $2m$-qubit random Clifford gate $U_3$ on $B$, which extends the stabilizer generators acting nontrivially on $B$ across the entire $B$, as Fig~\ref{fig:4blocks}(b) depicts. Finally, we completely depolarize $B_2$, which consists of $2mp$ of the qubits in $B$, assuming $p<1/2$. We denote the other $2m(1-p)$ qubits of $B$ as $B_1$. Due to the random Clifford gate $U_3$, the choice of $B_2$ from $B$ does not affect the result, and we select $B_2$ as the last $2mp$ qubits.

Now, we apply the complete depolarizations of the qubits in $B_2$ one by one, starting with the last qubit in $B_2$ corresponding to the $3m$-th column. We divide the tableau into two sections—the first $2m$ rows and the others. In the first section, we find two independent Pauli operators on the $3m$-th column—this is possible because we assume $m$ is very large. Without loss of generality, we find two rows with $X$ and $Z$, respectively, on the $3m$-th column. By permuting rows, we bring them to the first two rows. Using these two rows for row reduction, we ensure every row of the first section, except for the first two rows, has $\identity$ on the $3m$-th column. We repeat this procedure on the other section. As a result, we obtain a stabilizer tableau depicted in Fig.~\ref{fig:4blocks}(c). Then, we multiply the first and second rows of the second section by the first and second rows of the first section, respectively. As a result, the $3m$-th column shows only $\identity$'s except for the first two rows, as seen in Fig.~\ref{fig:4blocks}(d). Finally, eliminating the first two rows completes the complete depolarization algorithm introduced in Sec~\ref{subsec:stabilizer}, resulting in the stabilizer tableau in Fig.~\ref{fig:4blocks}(e). Consequently, the complete depolarization of a qubit in $B_2$ reduces the stabilizer tableau by two rows, yet preserves the number of rows randomly acting on $A$. The number of rows acting on $C$ randomly remains unchanged as well. Repeating this procedure on the remaining qubits in $B_2$ yields the stabilizer tableau of Fig.~\ref{fig:4blocks_clipped}(a).

For calculating $I(A:BC)$, $I(A:B)$, and $I(A:C|B)$, we take the clipped gauge. Note that the submatrix of the first $2m(1-2p)$ rows and columns of $A$ and $B_1$ forms a random stabilizer state. Similarly, the submatrix of the last $2m(1-2p)$ rows and columns of $B_1$ and $C$ also forms a random stabilizer state. Applying the maximal scrambling ansatz to these submatrices produces the stabilizer tableau in Fig.~\ref{fig:4blocks_clipped}(b). Through row reduction using the first $2m(1-2p)$ rows, we convert the first $m(1-2p)$ columns of the $4mp$ middle rows to all $\identity$'s. Similarly, using the last $2m(1-2p)$ rows, we convert the last $m(1-2p)$ columns of the $4mp$ middle rows to all $\identity$'s, resulting in the stabilizer tableau in Fig.~\ref{fig:4blocks_clipped}(c). Finally, by applying the maximal scrambling ansatz to the submatrix of the $4mp$ middle rows, we obtain the stabilizer tableau in Fig.~\ref{fig:4blocks_clipped}(d). This series of steps, assuming maximal scrambling ansatz for the submatrices, results in the clipped gauge for the entire tableau.

Given that the stabilizer tableau in Fig.~\ref{fig:4blocks_clipped}(d) is set in the clipped gauge, Eqs.~\eqref{eq:Clipped_MI} and \eqref{eq:Clipped_CMI} are applicable for calculating $I(A:B)$, $I(A:BC)$, and $I(A:C|B)$. Since $A$ and $B$ are adjacent, $I(A:B)$ is determined by the number of rows with left endpoints in $A$ and right endpoints in $B$,
\begin{equation}
    I(A:B) = 2m(1-2p).
\end{equation}
Likewise, one can see that
\begin{equation}
    I(A:BC) = 2m,
\end{equation}
which aligns with Eq.~(6) in the main text. Subsequently, applying the chain rule of mutual information,
\begin{equation}
    I(A:C|B) = 4mp.
\end{equation}
Since $A$ and $C$ are lightcone-separated, local noise combined with random Clifford gates facilitates the rapid spreading of CMI. This phenomenon can be interpreted as selective removal of correlation due to local noise. Before depolarizing the qubits in $B_2$, both $I(A:B)$, indicating short-range correlation, and $I(A:BC)$, indicating long-range correlation, were $4m$. Following the complete depolarization of $B_2$, only short-range correlation is diminished, while long-range correlation remains unchanged. This discrepancy between short-range and long-range correlations leads to the rapid spread of CMI.

\subsection{Coarse-grained circuit}

\begin{figure}
\includegraphics[width=\textwidth]{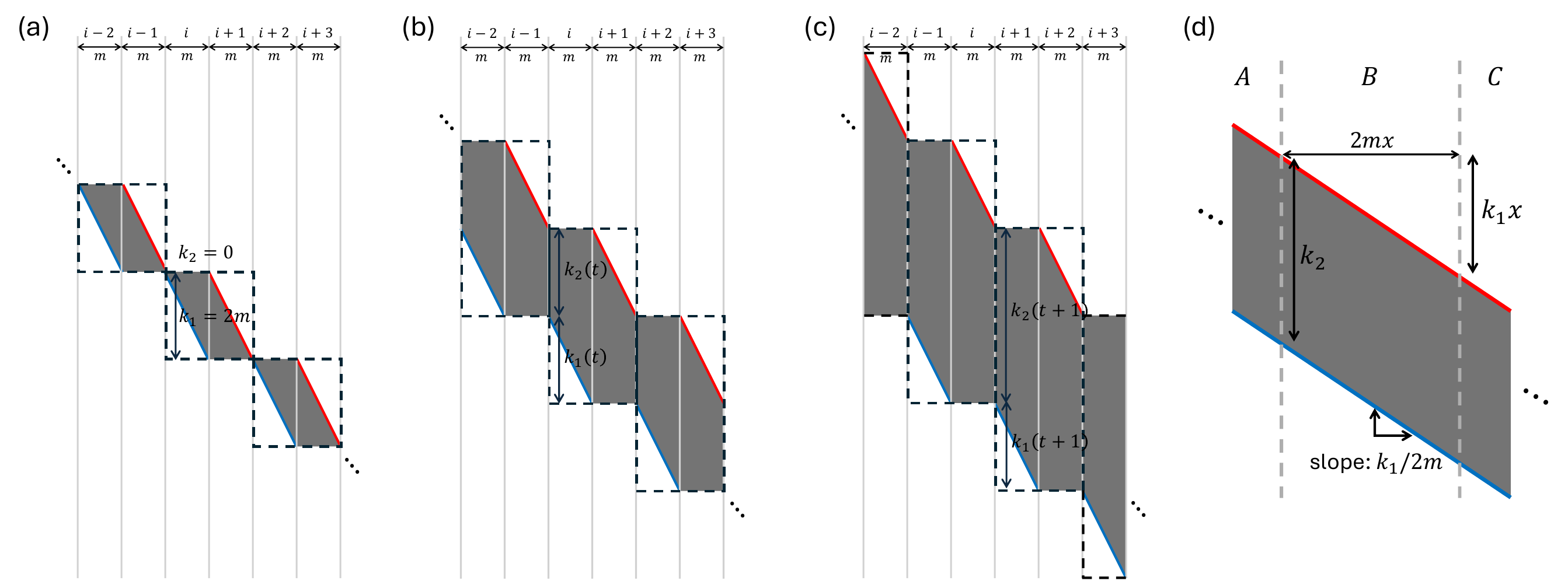}
\caption{\label{fig:coarse_noiseless}(a) Stabilizer tableau after applying the first layer of the random Clifford gates on the even pairs of blocks. The box of each pair of blocks is denoted by black dashed lines. Here, $k_1 = 2m$ and $k_2 = 0$. The maximal scrambling ansatz is applied to each box. (b) Stabilizer tableau at the timestep $t$, with $k_1 = k_1(t)$ and $k_2 = k_2(t)$ (c) The stabilizer tableau at the timestep $t+1$. Because of the layer of random Clifford gates at $t+1$, the height of each box increases. (d) Stabilizer tableau after omitting the microscopic details. It consists of a diagonal stripe with the slope of $k_1/2m$ and the width of $k_2$.}
\centering
\end{figure}

Here, we generalize the method used to analyze the four-block example to derive the analytical results for the coarse-grained circuit, considering the limit of a large coarse-graining factor. The coarse-grained random circuit of $N$ blocks of $m$ qubits, depicted in Fig.~2 in the main text, begins with the initial product state $\ket{0}^{\otimes mN}$. Here, $m$ is the coarse-graining factor that controls the local dimensionality. With $N$ required to be even and significantly large, we apply $2m$-qubit random Clifford gates on the even pairs of blocks $(1,2), (3,4), \cdots, (N-1,N)$ at even time steps $t=0, 2, \cdots$, and on the odd pairs $(2,3), (4,5), \cdots, (N-2,N-1)$ at odd time steps $t=1, 3, \cdots$. Following each random gate, we apply a heralded depolarizing channel on each qubit with an error rate $p$. This channel completely depolarizes the qubit with probability $p$, which controls the noise strength in the circuit. At every timestep, we divide the system into three regions: $A$, $B$, and $C$, as the first $N/2-x$ blocks, the middle $2x$ blocks, and the last $N/2-x$ blocks, respectively. Then, we calculate the normalized CMI, $I^{\mathrm{norm}}(A:C|B)$, which is defined as $I(A:C|B)/m$.

Now, assuming $N \rightarrow \infty$ and $m \rightarrow \infty$, we derive Eqs.~(8--10) in the main text, starting with the noiseless case where $p=0$. The initial state $\ket{0}^{\otimes mN}$ corresponds to a stabilizer tableau with $Z$'s on its diagonal and $\identity$'s elsewhere. For the stabilizer tableau at even (or odd) time steps, we assign a \textit{box} for each even (or odd) pair of blocks. A box of the $i$-th and $(i+1)$-th blocks is a submatrix comprising columns of those two blocks and rows acting nontrivially on one of the blocks. We denote the indices of the first and last rows in the box as $T_{i,i+1}$ and $B_{i,i+1}$, respectively. Additionally, we denote the number of rows with the left endpoint inside the box as $k_1$ and the number of the other rows as $k_2$. In the large $N$ limit, the stabilizer tableau exhibits a repeating pattern of boxes due to this translational invariance. Therefore, the indices of blocks can be omitted when denoting $k_1$ and $k_2$. These notations allow us to show that,
\begin{align}
    k_1 &= B_{i-2,i-1} - B_{i,i+1},\\
    k_2 &= T_{i,i+1} - B_{i-2,i-1} + 1,
\end{align}
for $i=1,3,5,\dots$ (or $i=2,4,6,\dots$) at even (or odd) timesteps. For instance, in the tableau of the initial state, left and right endpoints are located on the diagonal, yielding $T_{i,i+1}=mi$ and $B_{i,i+1}=m(i+1)$ for $i=1, 3, 5, \dots$, leading to $k_1=2m$ and $k_2 = 0$. The first layer of $2m$-qubit random Clifford gates fills all boxes with random stabilizers. We apply the maximal scrambling ansatz to each box, resulting in Fig.~\ref{fig:coarse_noiseless}(a). Notably, adopting the clipped gauge for each box results in applying the clipped gauge to the entire stabilizer tableau.

To observe the stabilizer tableau's evolution, consider the tableau at timestep $t$ with $k_1 = k_1(t)$ and $k_2 = k_2(t)$, as depicted in Fig.~\ref{fig:coarse_noiseless}(b). Suppose the box of the $i$-th and $(i+1)$-th blocks has $T_{i,i+1} = T_{i,i+1}(t)$ and $B_{i,i+1} = B_{i,i+1}(t)$. Then, at the next timestep, the box of the $(i+1)$-th and $(i+2)$-th blocks will have $T_{i+1,i+2}(t+1) = T_{i,i+1}(t)$ and $B_{i+1,i+2}(t+1) = B_{i+2,i+3}(t)$. Therefore, as Figs.~\ref{fig:coarse_noiseless}(b) and (c) illustrate, $k_1(t+1) = k_1(t)$ and $k_2(t+1) = k_1(t) + k_2(t)$. The application of a layer of random Clifford gates at timestep $t+1$ fills each box with the random stabilizers. Then, subsequently taking the clipped gauge results in the tableau seen in Fig.~\ref{fig:coarse_noiseless}(c). Consequently, the height of each box increases as $k_2$ grows while $k_1$ remains unchanged. The following equations summarize how the stabilizer tableau evolves:
\begin{align}
    \frac{dk_1}{dt} &= 0, \label{eq:diffeq1-1} \\
    \frac{dk_2}{dt} &= k_1, \label{eq:diffeq1-2}
\end{align}
initiating with $k_1(t=0) = 2m$ and $k_2(t=0) = 0$. Therefore,
\begin{align}
    k_1 &= 2m,\\
    k_2 &= 2mt.
\end{align}
Given $x\gg 1$, the ``stair structure" of the stabilizer tableau becomes negligible. By abstracting away those microscopic details, we obtain a simplified representation of the stabilizer tableau, depicted in Fig.~\ref{fig:coarse_noiseless}(d), with a diagonal stripe characterized by a slope of $k_1/2m$ and a vertical width of $k_2$. Then, the calculation of $I(A:B)$ and $I(A:BC)$ is straightforward by leveraging the clipped gauge's properties. By Eq.~\eqref{eq:Clipped_MI}, $I(A:B)$ corresponds to the number of rows with left endpoints in $A$ and right endpoints in $B$. With $2mx$ qubits in $B$,
\begin{equation}\label{eq:IAB_k}
    I(A:B) = \min\{k_1x, k_2\}.
\end{equation}
Similarly, as $I(A:BC)$ represents the number of rows with left endpoints in $A$ and right endpoints in $B$ or $C$,
\begin{equation}\label{eq:IABC_k}
    I(A:BC) = k_2.
\end{equation}
Thus, we derive Eqs.~(8--10) in the main text for the noiseless scenario ($p=0$),
\begin{align}
    I(A:BC) &= 2mt,\\
    I(A:B) &= \min\{2mx, I(A:BC)\},\\
    I(A:C|B) &= \max\{2mt - 2mx,0\}.
\end{align}

\begin{figure}
\includegraphics[width=\textwidth]{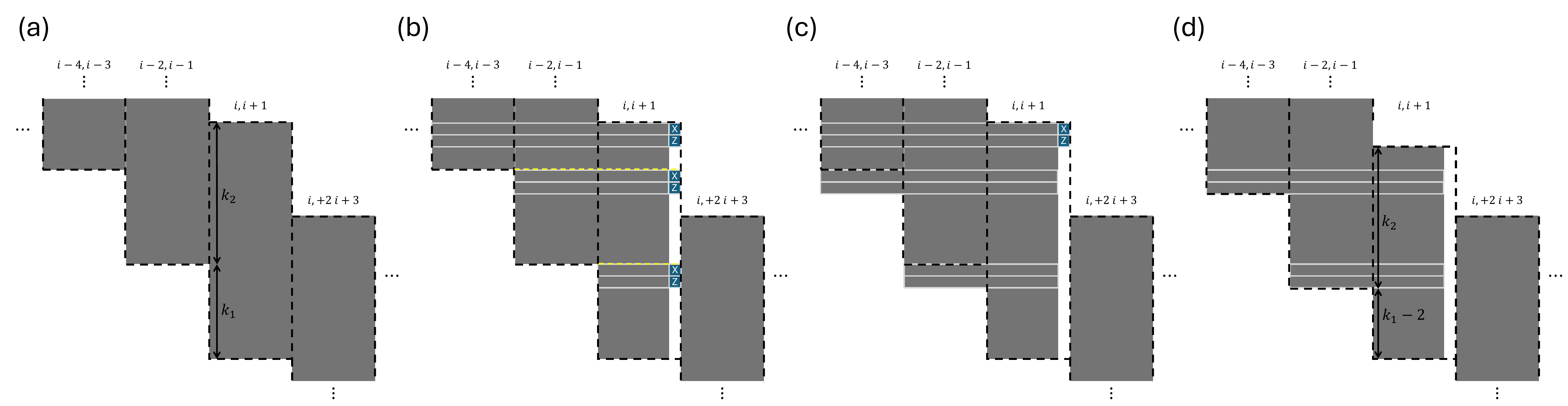}
\caption{\label{fig:coarse_traceout}Procedure of completely depolarizing the last qubit of the box of the $i$-th and $(i+1)$-th blocks. (a) The stabilizer tableau where each row's support is across at most three boxes. (b) We divide the box of the $i$-th and $(i+1)$-th blocks into three sections, as marked with the yellow dashed lines. On the last column of the box, we find two independent Puuali operators in each section and bring them to the first and second rows. By doing row-reduction, we make the other rows all $\identity$'s on the last column of the box. (c) The stabilizer tableau after we multiply the first two rows of the last section by the first two rows of the second section, respectively, and then multiply the first two rows of the second section by the first two rows of the first section, respectively. (d) Removing the first two rows of the box completes the complete depolarization on the last qubit of the box. This results in the reduction of $k_1$ by $2$ while $k_2$ is unchanged.}
\centering
\end{figure}

Now, we analyze the noisy case of $p>0$. As we introduce heralded depolarizing channels, Eqs.~\eqref{eq:diffeq1-1} and \eqref{eq:diffeq1-2} need to be modified accordingly. A key observation is that local depolarization decreases $k_1$, but not $k_2$, as long as $t<1/2p$. This can be demonstrated by extending the method used in the four-block example. However, unlike the four-block example, each row's support can extend across more than two boxes. Here, we demonstrate that local depolarization decreases $k_1$ only when each row's support spans at most three boxes, as illustrated in Fig.~\ref{fig:coarse_traceout}(a). Generalizing this to the cases where each row's support spans more than three boxes is straightforward.

Assuming the tableau is partitioned into boxes characterized by $k_1$ and $k_2$ before applying the clipped gauge, we first completely depolarize a single qubit in the box of the $i$-th and $(i+1)$-th blocks. Without loss of generality, we choose the box's last qubit for depolarization to observe its impact on $k_1$ and $k_2$. The box is divided into three sections: rows $T_{i,i+1}, T_{i,i+1}+1, \dots, B_{i-4,i-3}$; rows $B_{i-4,i-3}+1, B_{i-4,i-3}+2, \dots, B_{i-2,i-1}$; and rows $B_{i-2,i-1}+1, B_{i-2,i-1}+2, \dots, B_{i,i+1}$. Similar to the four-block example, we find two independent Pauli operators on the box's last column within the first section. This is possible because of the assumption that $m \gg 1$. We locate two rows with $X$ and $Z$, respectively, on the box's last column. By permuting rows, these are brought to the forefront in the first section without altering the tableau's configuration. Subsequent row reductions render all other rows in the first section as $\identity$ on the last column. This procedure is repeated for the other sections, resulting in the stabilizer tableau shown in Fig.~\ref{fig:coarse_traceout}(b). After that, we multiply the first two rows of the last section by the first two rows of the second section, respectively. Again, we multiply the first two rows of the second section by the first two rows of the first section, respectively, and we get the stabilizer tableau in Fig.~\ref{fig:coarse_traceout}(c). Finally, deleting the first two rows of the box completes the algorithm of complete depolarization, resulting in the stabilizer tableau in Fig.~\ref{fig:coarse_traceout}(d). This process removes two rows from the concerned box, reducing $k_1$ by 2 while $k_2$ remains unchanged. The configurations of the other boxes stay unaltered, indicating that depolarization within a specific box does not impact the other boxes. Consequently, depolarizing a single qubit in every box diminishes $k_1$ by $2$, maintaining $k_2$ constant, assuming $k_1 \ge 0$.

In the coarse-grained circuit, each timestep involves depolarizing $mp$ qubits per block, equating to $2mp$ qubits per box. As a result, $k_1$ is diminished by $4mp$ at every timestep until becoming zero. This modifies the Eqs.~\eqref{eq:diffeq1-1} and \eqref{eq:diffeq1-2} as follows:
\begin{align}
    \frac{dk_1}{dt} &= -4mp, \\
    \frac{dk_2}{dt} &= k_1,
\end{align}
before $k_1$ becomes zero. Consequently, we derive:
\begin{align}
    k_1(t) &= 2m(1-2pt),\\
    k_2(t) &= 2mt(1-pt),
\end{align}
up until the critical timestep $t_c=1/2p$. Employing Eqs.~\eqref{eq:IAB_k} and \eqref{eq:IABC_k}, we drive Eqs.~(8--10) in the main text for the noisy circuit of $p>0$,
\begin{align}
    I(A:BC) &= 2mt(1-pt),\\
    I(A:B) &= \min\{2m(1-2pt)x, I(A:BC)\},\\
    I(A:C|B) &= \max\{2mt(1-pt) - 2m(1-2pt)x,0\},
\end{align}
for $t < t_c$. At $t=t_c$, $k_1$ becomes zero, and $k_2=m/2p$. This corresponds to the stabilizer tableau of $m/2p$ rows, where each row acts randomly on the entire system that consists of $Nm$ qubits. In the next timestep, we apply complete depolarization on $\simeq Nmp$ qubits. Since we take the large $N$ limit, all rows of the stabilizer tableau are removed, and thus, the state becomes maximally mixed. Therefore,
\begin{equation}\label{eq:anal_main}
    I^{\mathrm{norm}}(A:C|B) =
    \begin{cases}
        \max\{ 2t(1-pt) - 2(1-2pt)x , 0 \}, & t \le t_c \\
        0. & t > t_c
    \end{cases}
\end{equation}

\section{\label{sec:detailed_results}Detailed numerical results}

\begin{figure}
\includegraphics[width=\textwidth]{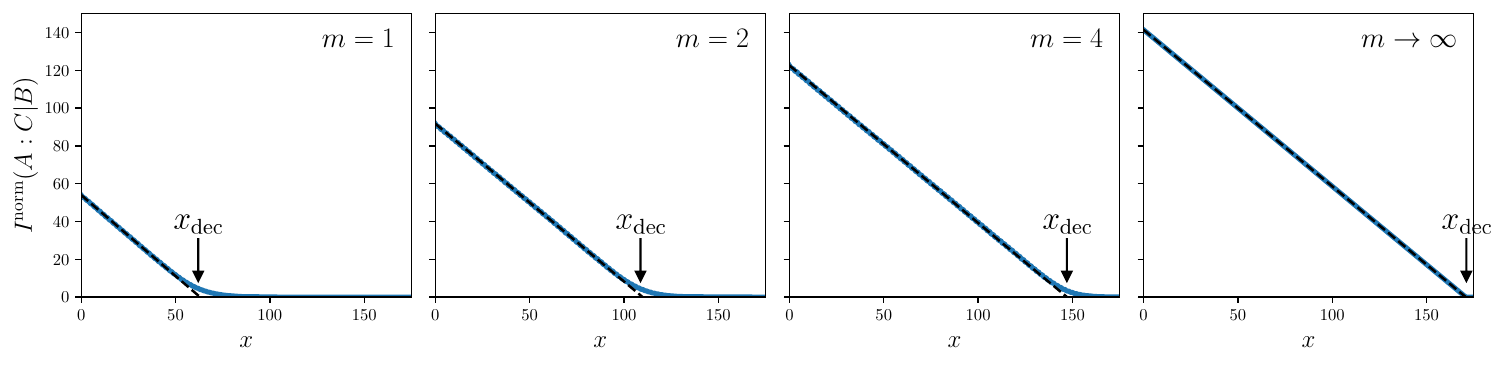}
\caption{\label{fig:SM_decay}Decay of $I^{\mathrm{norm}}(A:C|B)$ in various coarse-graining factors at timestep $t=100$. The first three panels are from the numerical results with $m= 1, 2, 4$, respectively. All numerical simulations are performed for the circuits with $N=2^{10}$ and $p=1.5\times 2^{-9}$, averaged over $1000$ circuit realizations. The last panel is an analytical result of \eqref{eq:anal_main} with $m \rightarrow \infty$ and $p=1.5 \times 2^{-9}$. Unlike the analytical results, $I^{\mathrm{norm}}(A:C|B)$ has an exponential tail after the initial decay in the numerical results with the finite coarse-graining factor. Black dashed lines are the linear fittings of the initial linear decay of $I^{\mathrm{norm}}(A:C|B)$, and $x_{\mathrm{dec}}$'s are defined as the $x$-intercept of the fitted lines.}
\centering
\end{figure}

\begin{figure}
\includegraphics[width=\textwidth]{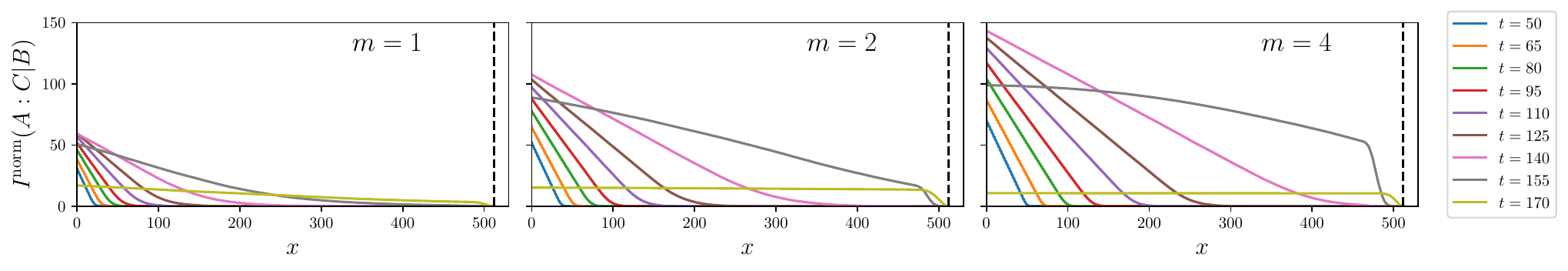}
\caption{\label{fig:SM_decay_boundary}Decay of $I^{\mathrm{norm}}(A:C|B)$ at various timesteps with the coarse-graining factors $1,2,4$, respectively. All numerical simulations are performed for the circuits with $N=2^{10}$ and $p=1.0\times 2^{-9}$, averaged over $1000$ circuit realizations. Black dashed lines indicate the boundary $x=N/2$. As time evolves, CMI propagates until it meets the boundary, and then the decay profile starts to be distorted.}
\centering
\end{figure}

\begin{figure}
\includegraphics[width=\textwidth]{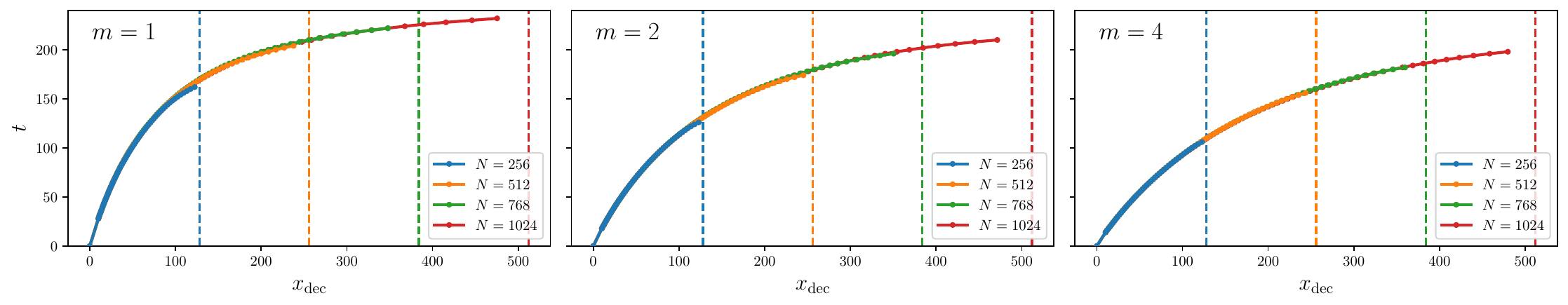}
\caption{\label{fig:SM_boundary_change_N}Propagation of $x_\mathrm{dec}$ with the coarse-graining factors $1,2,4$. All numerical simulations are performed for $p=1.0\times 2^{-9}$ with various values of $N$, averaged over $1000$ circuit realizations. Dashed lines indicate the boundary $x=N/2$. As we omitted the distortion from the boundary, propagation of $x_{\mathrm{dec}}$ does not depend on $N$.}
\centering
\end{figure}

This section presents detailed numerical results, elucidating the discrepancies between analytical results and numerical simulations due to finite coarse-graining factors $m$ and the number of blocks $N$. When deriving the analytical result of Eq.~\eqref{eq:anal_main}, we assumed that both $m$ and $N$ are indefinitely large. However, in numerical simulations, both $m$ and $N$ are necessarily finite, leading to differences between the analytical and numerical results. Specifically, a finite $m$ results in a deviation from the linear decay of CMI with respect to $x$, which appears in Eq.~\eqref{eq:anal_main}, while a finite $N$ introduces boundary effects.

In the theoretical model assuming $m, N \gg 1$, CMI decays linearly in terms of $x$ until it reaches zero. In contrast, for circuits with a finite $m$, Fig.~\ref{fig:SM_decay} reveals that CMI exhibits an exponential tail following the initial linear decay. This is because, although exponentially unlikely in a random circuit, there exists a constant depth quantum circuit that spreads CMI for an unbounded distance~\cite{zou_spurious_2016, williamson_spurious_2019, kato_toy_2020, kim_universal_2023}. This exponential tail diminishes with an increase in the coarse-graining factor $m$ and disappears in the limit as $m \rightarrow \infty$.

In the main text, for analytical results with $m \rightarrow \infty$, we define $x_{\mathrm{dec}}$ as the value of $x$ at which $I^{\mathrm{norm}}(A:C|B)$ becomes zero. However, due to the presence of the exponential tail in the decay, this definition of $x_{\mathrm{dec}}$ cannot be directly applied to numerical results with finite coarse-graining factors. To address this, we determine $x_{\mathrm{dec}}$ based on the initial linear decay. More specifically, linear fitting is performed on sections exhibiting linear decay, and the $x$-intercept of the fitted line is selected as $x_{\mathrm{dec}}$, as illustrated in Fig.~\ref{fig:SM_decay}.

The finite value of $N$ also creates a discrepancy between the analytical results and the numerical results, as there are a finite number of blocks of qubits. Fig.~\ref{fig:SM_decay_boundary} shows the effect of the boundary introduced by finite $N$. As CMI spreads, it eventually encounters the system's boundaries. Initially, well before reaching the boundary, CMI decays linearly, followed by an exponential tail. Once the spreading meets the boundary, the decay profile starts to be distorted. To avoid the impact of boundary-induced distortions, $x_{\mathrm{dec}}$ is calculated exclusively for timesteps prior to the distortion from the boundary. This approach isolates the bulk dynamics from the boundary effect. As demonstrated in Fig.~\ref{fig:SM_boundary_change_N}, the propagation of $x_{\mathrm{dec}}$ appears independent of $N$, confirming that the remaining part of $x_{\mathrm{dec}}$ is unaffected by the boundary.

\section{\label{entanglement}Generating entanglement between a large distance}

\begin{figure}
\includegraphics[width=\textwidth]{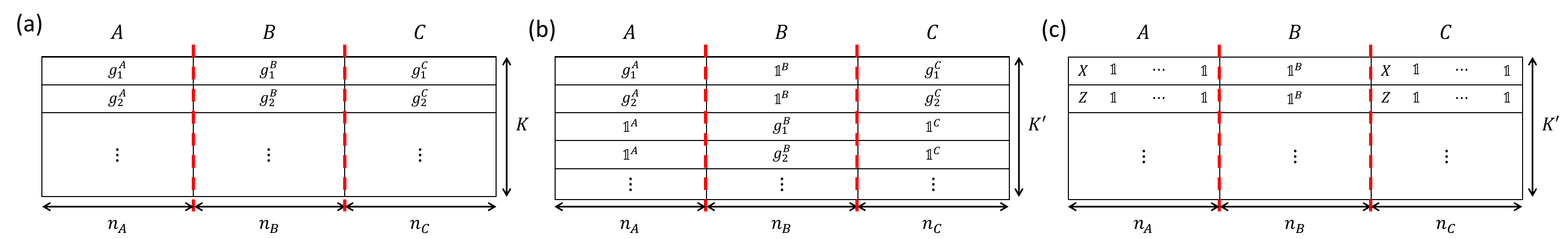}
\caption{\label{fig:long_entanglement}(a) Stabilizer tableau of a random stabilizer state with $(n_A + n_B + n_C)$ qubits and $K$ stabilizer generators. We divide the system into $A$, $B$, and $C$, which are the first $n_A$ qubits, the second $n_B$ qubits, and the last $n_C$ qubits, respectively. Here, we choose $g_1$ and $g_2$ that satisfy the anticommutation conditions of Eqs.~\eqref{eq:gA} and \eqref{eq:gC}. (b) Stabilizer tableau of the state after measuring with $g_1^B$ and $g_2^B$. Because of those measurements, the number of generators can increase. (c) Stabilizer tableau after the additional local operations on $A$ and $C$, where $A$ and $C$ share a Bell pair.}
\centering
\end{figure}

\begin{figure}
\includegraphics[width=\textwidth]{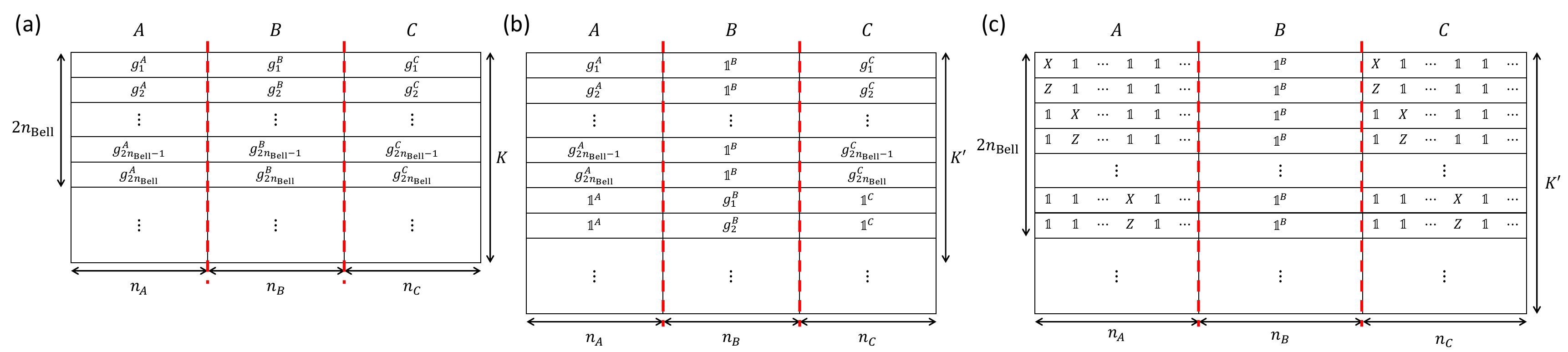}
\caption{\label{fig:long_entanglement_2}(a) Stabilizer tableau of the state where there are $n_\mathrm{Bell}$ pairs of stabilizer generators $(g_1, g_2), (g_3, g_4), \dots, (g_{n_\mathrm{Bell}-1}, g_{n_\mathrm{Bell}})$ where each generator among these pairs anticommutes with the paired generator for the parts acting on $A$ or $C$ and commutes with the others, and all $g^B_i$'s commutes with each other. (b) Stabilizer tableau of the state after measuring $B$ with $g_1^B, g_2^B, \dots, g_{2n_\mathrm{Bell}}^B$. (c) Stabilizer tableau after the additional local operations on $A$ and $C$, where $A$ and $C$ share $n_{\mathrm{Bell}}$ Bell pairs.}
\centering
\end{figure}

The divergence of $x_{\mathrm{dec}}$ at the critical timestep $t_c$ implies CMI spreads throughout the entire system. This section demonstrates that this pervasive conditional dependence is not sorely composed of classical correlations but the multipartite quantum correlation extending throughout the entire system. To see this, we show that the regions $A$ and $C$, separated by an arbitrarily large region $B$, can achieve quantum correlation through appropriate measurements on $B$.

In Sec.\ref{sec:coarse-graining}, we discussed the divergence of $x_{\mathrm{dec}}$ attributes to every stabilizer generator spanning the entire system in the clipped gauge. We consider a scenario where $K$ stabilizer generators are left at $t_c$. Given these stabilizers extend across the entire system, we assume the state to be a random stabilizer state, selecting $g_1, g_2, \dots, g_K$ as independent Pauli strings that are commuting. The system is partitioned into $A$, $B$, and $C$—the first $n_A$ qubits, the second $n_B$ qubits, and the last $n_C$ qubits, respectively, and we write,
\begin{align}
    g_i &= g_i^A \otimes g_i^B \otimes g_i^C, \\
\end{align}
for $i=1,2,\dots ,K$. The stabilizer tableau is depicted in Fig.~\ref{fig:long_entanglement}(a). Here, $n_A$, $n_B$, and $n_C$ are considered indefinitely large, indicating an extensive separation between $A$ and $C$.

Among those $K$ stabilizer generators, we choose two generators, $g_1$ and $g_2$, such that
\begin{align}
    g_1^A g_2^A &= - g_2^A g_1^A,\label{eq:gA}\\
    g_1^C g_2^C &= - g_2^C g_1^C.\label{eq:gC}
\end{align}
Given that $K$ is sufficiently large, we can choose such pairs of generators with high probability. To see this, recall that for a random stabilizer state, the stabilizer generators $g_1, g_2, \dots, g_K$ are random Pauli strings under the constraints of $[g_i, g_j] = 0$ for $i,j = 1,2,\dots, K$ and independence of them. However, focusing solely on the submatrix encompassing the columns for $A$ and $C$, one can expect that this submatrix is insensitive to these global constraints because the significant number of columns, $n_B$, are neglected in Eqs.~\eqref{eq:gA} and \eqref{eq:gC}. Thus, we assume that $g_i^A$ and $g_i^B$ for $i=1,2,\dots K$ are chosen as random Pauli strings without any constraints. Therefore, for any pairs of stabilizer generators $g_i$ and $g_j$, given that non-trivial Pauli strings anti-commute with half of the elements in the Pauli group, the probability that the conditions Eqs.~\eqref{eq:gA} and \eqref{eq:gC} are fulfilled is:
\begin{equation}
    \frac{1}{4}\left(1-4^{-n_A}\right)\left(1-4^{-n_C}\right) \simeq \frac{1}{4},
\end{equation}
where factors $(1-4^{-n_A})$ and $(1-4^{-n_C})$ are for excluding the cases that we choose $g_1^A = \identity$ or $g_1^C = \identity$. This shows that there exists a pair of generators $g_1$ and $g_2$ satisfying the conditions Eqs.~\eqref{eq:gA} and \eqref{eq:gC} with the probability of $\simeq 1-(3/4)^{K(K-1)/2}$. Consequently, it is exponentially likely for us to be able to choose such $g_1$ and $g_2$.

Here, we claim that conducting measurements on $B$ with the observables $g_1^B$ and $g_2^B$ can effectively generate a Bell pair between $A$ and $C$. First, the anticommutation conditions of Eqs.~\eqref{eq:gA} and \eqref{eq:gC}, alongside the constraint $[g_1,g_2]=0$, guarantee that
\begin{equation}\label{eq:gB}
    [g_1^B, g_2^B] =0, 
\end{equation}
so $g_1^B$ and $g_2^B$ are compatible observables. As mentioned in Sec.~\ref{subsec:stabilizer}, the measurement outcomes only determine the overall phases of the stabilizer generators, and thus, we assume that the measurement outcomes are $+1$ without loss of generality. Then, the measurements make the generating set become $\{g_1, g_2, g_1^B, g_2^B, \dots \}$, possibly increasing the number of generators. We can further multiply $g_1$ and $g_2$ by $g_1^B$ and $g_2^B$ respectively, resulting in the stabilizer tableau in Fig.~\ref{fig:long_entanglement}(b). Finally, by applying local Clifford gates on $A$ and $C$, we can convert the first two rows of the tableau to $X_{1}X_{n_{A}+n_{B}+1}$ and $Z_{1}Z_{n_{A}+n_{B}+1}$, respectively, as seen in Fig.~\ref{fig:long_entanglement}(c)~\cite{audenaert_entanglement_2005}. Here, $X_i$ and $Z_i$ stand for $X$ and $Z$ acting on $i$-th qubit, respectively. Importantly, for the stabilizer state corresponding to Fig.~\ref{fig:long_entanglement}(c), $A$ and $C$ share a Bell pair. In summary, at the critical timestep, the local observers in $A$ and $C$ can share a quantum entanglement by doing measurements on $B$ and informing the measurement outcomes to $A$ and $C$.

We can generalize this protocol to generate more than one Bell pair. Suppose we find $2n_\mathrm{Bell}$ generators, $g_1, g_2, \dots, g_{2n_\mathrm{Bell}-1}, g_{2n_\mathrm{Bell}}$ such that
\begin{align}
    g^A_{i}g^A_{i+1} &= - g^A_{i+1}g^A_{i},\\
    g^C_{i}g^C_{i+1} &= - g^C_{i+1}g^C_{i},
\end{align}
for $i=1,3, \dots, 2n_\mathrm{Bell}-1$, and
\begin{align}
    [g^A_{i},g^A_{j}] &= 0,\\
    [g^C_{i},g^C_{j}] &= 0,
\end{align}
for $i, j = 1,2,\dots, 2n_\mathrm{Bell}$ such that $\{i,j\} \notin \{ \{1,2\},\{3,4\}, \dots \{2n_\mathrm{Bell}-1,2n_\mathrm{Bell}\}\}$, and lastly,
\begin{align}
    [g^B_{i},g^B_{i+1}] &= 0,\\
    [g^B_{i},g^B_{i+1}] &= 0,
\end{align}
for all $i, j = 1,2,\dots, 2n_\mathrm{Bell}$. In other words, there are $n_\mathrm{Bell}$ pairs of generators where each generator among those pairs anticommutes with the paired generator for the parts acting on $A$ or $C$ and commutes with the others, and all $g^B_i$'s commutes with each other. Fig.~\ref{fig:long_entanglement_2}(a) depicts the corresponding stabilizer tableau. Then, in the same way as above, measuring $B$ with $g^B_{1}, g^B_{2}, \dots, g^B_{2n_\mathrm{Bell}}$ makes the stabilizer tableau in the form of Fig.~\ref{fig:long_entanglement_2}(b), and we can distill $n_\mathrm{Bell}$ Bell pairs between $A$ and $C$ by applying local operations on $A$ and $C$ and thus achieving a stabilizer tableau in Fig.~\ref{fig:long_entanglement_2}(c).

Clearly, the number of Bell pairs that are induced by the measurements on $B$ is bounded by $K/2$, as seen in Fig~\ref{fig:long_entanglement_2}(a). Meanwhile, at the critical timestep $t_c$, every stabilizer generator extends across the entire system in the clipped gauge, resulting in $I(A:C|B) = K$. Therefore, for the number of distillable Bell pairs $n_\mathrm{Bell}$,
\begin{equation}
    n_\mathrm{Bell} \le I(A:C|B)/2,
\end{equation}
at the critical timestep $t_c$. In other words, CMI acts as an upper bound of the number of distillable Bell pairs when $t=t_c$. Meanwhile, as noted above, it is exponentially unlikely that there is no distillable Bell pair. Therefore, the lower bound of the expected number of Bell pairs is at least constant.

\section{Generalization to Haar random circuits with depolarizing channel}

To leverage the stabilizer formalism for theoretical analysis and large-scale numerical simulations, we employed Clifford gates as the random elements and the heralded depolarizing channel as the decoherence model in our coarse-grained random circuit. We expect that the conclusions drawn from Random Clifford circuits with heralded depolarizing channels are generic and can be extended to Haar random circuits with depolarizing channels. This section elaborates on the distinctions between depolarizing and heralded depolarizing channels and presents numerical evidence supporting the generalization of our findings to Haar random circuits.

\subsection{\label{subsec:depo_vs_heralded_depo}Depolarizing channel vs. heralded depolarizing channel}

First, we give a detailed discussion on the heralded depolarizing channel, focusing on its difference from the depolarizing channel. The depolarizing channel on the qubit $i$ with an error rate $p$ is defined as
\begin{equation}\label{eq:depo}
    \rho \rightarrow (1-p)\rho + p \Tr_{i}\rho \otimes \identity_{i}/2.
\end{equation}
This noise channel describes the circumstance in which an error completely removes information in the qubit with a probability of $p$. Furthermore, since the output state is a mixture of the input state and the maximally mixed state, it does not reveal whether the error happened or not. On the other hand, the heralded depolarizing channel with an error rate $p$ is defined as
\begin{equation}\label{eq:hdepo1}
\rho \mapsto
\begin{cases}
    \Tr_{i} \rho \otimes \identity_{i}/2 &\text{with Prob. of $p$,}\\
    \rho &\text{with Prob. of $1-p$,}
\end{cases}
\end{equation}
The heralded depolarizing channel also describes losing the information in the qubit with a probability of $p$. However, as opposed to the depolarizing channel, the output state is either the same as the input or maximally mixed state, but not a mixture of the two cases. Specifically, consider a coarse-grained random circuit with $N$ blocks of $m$ qubit and a depth $t$. Then, there are $Nmt$ spacetime locations where we apply heralded depolarizing channels. We assign $0$ and $1$ to those spacetime locations as $r \in \{0,1\}^{Nmt}$, and let $\rho^r$ be the output state of the circuit such that we apply complete depolarization on the spacetime locations where $1$ is assigned. Therefore, $r$ indicates the configuration of the errors, and $\rho^r$ is the output state corresponding to the configuration. Then, the circuit with the heralded depolarizing channels outputs $\rho^r$ with the probability of $p(r) = p^{\abs{r}}(1-p)^{Nmt - \abs{r}}$, where $\abs{r}$ is the Hamming weight of $r$. However, if we replace the heralded depolarizing channels with depolarizing channels, the output state is $\rho^{\mathrm{depo}} = \sum_{r \in \{0,1\}^{Nmt}} p(r) \rho^r$.

In our study, we calculate CMI averaged over the circuit realizations. The circuit realizations depend on the choice of unitary gates and the spacetime locations of errors $x$, and here, we focus on taking the average over the locations of errors. First, for the circuit with the heralded depolarizing channels, we calculate the following quantity,
\begin{equation}\label{eq:CMI_hdepo}
    \mathbb{E}_{r} I(A:C|B)_{\rho^r} = \sum_{r \in \{0,1\}^{Nmt}} p(r)I(A:C|B)_{\rho^r},
\end{equation}
where $\mathbb{E}_{r}$ represents averaging over the locations of errors. Meanwhile, for the circuit with the depolarizing channels, we get
\begin{equation}\label{eq:CMI_depo}
    I(A:C|B)_{\rho^{\mathrm{depo}}}.
\end{equation}
Since CMI is not a linear function, $\mathbb{E}_{r} I(A:C|B)_{\rho^r}$ and $I(A:C|B)_{\rho^{\mathrm{depo}}}$ are different in general. Moreover, there is no obvious inequality between them because CMI is neither convex nor concave.

To analyze the subtle difference between the depolarizing channel that leads to Eq.~\eqref{eq:CMI_depo} and the heralded depolarizing channel that leads to Eq.~\eqref{eq:CMI_hdepo}, we introduce an auxiliary system $R$ to represent the information of error locations. The auxiliary system $R$ has an orthonormal basis $\ket{r}$ for $r \in \{0,1\}^{Nmt}$. Let $S$ be the qubits in the circuits, and consider a joint state of $SR$,
\begin{equation}
    \sigma_{SR} = \sum_{r \in \{0, 1\}^{Nmt}} p(r)\rho^{r} \otimes |r\rangle\langle r|_R.
\end{equation}
Here, $r$ stores the configuration of errors that outputs $\rho^{r}$. Note that if we trace out the auxiliary system $R$, the output state becomes that of the coarse-grained circuit with the depolarizing channels: $\rho^{\mathrm{depo}} = \Tr_{R} \sigma_{SR}$. Therefore,
\begin{equation}\label{eq:CMI_depo2}
    I(A:C|B)_{\rho^{\mathrm{depo}}} = I(A:C|B)_{\sigma_{SR}}.
\end{equation}
On the other hand, for the circuit with the heralded depolarizing channels, the von Neumann entropy of a subsystem $A$ averaged over the error configurations is
\begin{align}
    \mathbb{E}_{r} \left[S(A)_{\rho^{r}}\right] &= \sum_{r \in \{0, 1\}^{Nmt}} p(r) \left[- \Tr \rho^{r}_{A}\log \rho^{r}_{A}\right] \\
    &= S(A|R)_{\sigma_{SR}},
\end{align}
where $S(A|R)_{\sigma_{SR}} = S(AR)_{\sigma_{SR}} - S(R)_{\sigma_{SR}}$ is the conditional entropy. Since taking average $\mathbb{E}_{r}[\cdot]$ is linear,
\begin{align}
    \mathbb{E}_{r} \left[I(A:C|B)_{\rho^{r}_{S}}\right] &= S(AB|R)_{\rho_{SR}} + S(BC|R)_{\rho_{SR}} - S(ABC|R)_{\rho_{SR}} - S(B|R)_{\rho_{SR}}\\
    &= I(A:C|BR)_{\rho_{SR}}.\label{eq:CMI_hdepo2}
\end{align}
Eqs.~\eqref{eq:CMI_depo2} and \eqref{eq:CMI_hdepo2} clearly show that the distinction of the heralded depolarizing channel from the depolarizing channel is the extra conditioning on the auxiliary system $R$ that contains the information of the error configuration. In other words, unlike a depolarizing channel, the heralded depolarizing channel heralds the spacetime locations of the errors, or equivalently, the spacetime locations of the errors are determined for each circuit realization. Since additional conditioning can either increase or decrease CMI~\cite{cover_elements_2006}, CMI from the circuit with the heralded depolarizing channels is neither a lower nor upper bound of that from the circuit with the depolarizing channels.

\subsection{Matrix product operators}\label{subsec:mpo_intro}

As discussed above, there are subtle differences between the depolarizing channel and the heralded depolarizing channel. Nevertheless, we expect them to exhibit similar qualitative behaviors when combined with random circuits. Additionally, since the Clifford group forms a unitary 3-design~\cite{webb_clifford_2016, zhu_multiqubit_2017}, the averaged quantities, like averaged CMI, should closely resemble those obtained in Haar random circuits.

To validate our expectation that the spreading of CMI in Haar random circuits with depolarizing channels resembles the behavior seen in Clifford random circuits with heralded depolarizing channels, we perform the numerical simulation of the Haar random circuits with depolarizing channels. Simulating quantum states precisely requires an exponential number of parameters, making classical computation intractable for general quantum states. However, using the approach of matrix product state (MPS)~\cite{vidal_efficient_2003}, a subset of quantum systems with limited entanglement on a one-dimensional chain can be efficiently simulated. In this work, to simulate Haar random circuits with depolarizing channels, we use a matrix product operator (MPO) approach~\cite{verstraete_matrix_2004, zwolak_mixed-state_2004, noh_efficient_2020}, a mixed-state generalization of MPS. Specifically, we follow the method employed in Ref.~\cite{noh_efficient_2020}, and this section provides a brief overview of the MPO method and its application in our simulations.

Consider a density matrix of $N$ qudits, each of which has $d$ levels:
\begin{equation}
    \rho = \sum_{i_1, \dots, i_N, j_1, \dots, j_N=0}^{d-1} \rho_{i_1, j_1, \dots, i_N, j_N} \ketbra{i_1\dots i_N}{j_1 \dots j_N}.
\end{equation}
Here, the tensor $\rho_{i_1, j_1, \dots, i_N, j_N}$ contains exponentially many parameters. To circumvent this problem, we first vectorize the matrix as follows. For each qubit $n=1,\dots,N$, we map $\ketbra{i_n}{j_n}$ into a basis of a $d^2$-dimensional vector space,
$$
\ketbra{i_{n}}{j_{n}} \rightarrow |I_{n}\rrangle,
$$
where $I_{n} = di_{n}+j_{n}$. Thus, we convert the $d\times d$ matrix into a vector with $d^2$ elements for each $n$. As a result, we obtain a vectorized form of the density matrix:
\begin{equation}
    |\rho\rrangle = \sum_{I_1, \dots, I_N = 0}^{d^2-1} \rho_{I_1, \dots, I_N} |I_1, \dots, I_N\rrangle.
\end{equation}
Next, just as the MPS, we decompose the tensor $\rho_{I_1, \dots, I_N}$ as follows:
\begin{equation}\label{eq:MPO}
    \rho_{I_1, \dots, I_N} = \sum_{\alpha_1,\alpha_2,\alpha_3, \dots, \alpha_{N-1}}
    \Gamma^{[1]I_{1}}_{\alpha_{1}} \lambda^{[1]}_{\alpha_{1}} \Gamma^{[2]I_{2}}_{\alpha_{1},\alpha_{2}} \lambda^{[2]}_{\alpha_{2}} \Gamma^{[3]I_{3}}_{\alpha_{2},\alpha_{3}} \dots \lambda^{[N-1]}_{\alpha_{N-1}} \Gamma^{[N]I_{N}}_{\alpha_{N-1}}.
\end{equation}
This form achieves a Schmidt decomposition for the bipartition $[1\dots l]:[(l+1)\dots N]$, for all $l = 1,2,\dots N-1$,
\begin{equation}
    |\rho\rrangle = \sum_{\alpha_l} \lambda_{\alpha_l} |\Phi^{[1\dots l]}_{\alpha_l}\rrangle |\Phi^{[(l+1)\dots N]}_{\alpha_l}\rrangle
\end{equation}
where $\lambda^{[l]}_{\alpha_l}$'s are the singular values and the orthonormal bases are given by:
\begin{align}
    |\Phi^{[1\dots l]}_{\alpha_l}\rrangle &= 
    \sum_{I_1,\dots,I_l}\sum_{\alpha_1, \dots, \alpha_{l-1}} \Gamma^{[1]I_1}_{\alpha_1}\lambda^{[1]}_{\alpha_1}\dots \lambda^{[l-1]}_{\alpha_{l-1}}\Gamma^{[l]I_{l}}_{\alpha_{l-1}\alpha_{l}} |I_1 \dots I_l\rrangle,
    \\
    |\Phi^{[(l+1)\dots N]}_{\alpha_l}\rrangle &= 
    \sum_{I_{1+1},\dots,I_N}\sum_{\alpha_{l+1}, \dots, \alpha_{N-1}} \Gamma^{[l+1]I_{l+1}}_{\alpha_{l}\alpha_{l+1}}\lambda^{[l+1]}_{\alpha_{l+1}}\dots \lambda^{[N-1]}_{\alpha_{N-1}}\Gamma^{[N]I_{N}}_{\alpha_{N-1}} |I_{1+1} \dots I_N\rrangle.
\end{align}
Here, the singular values $\lambda^{[l]}_{0},\lambda^{[l]}_{1},\lambda^{[l]}_{2},\dots$ are arranged in a descending order. Unlike MPS, these singular values are not directly linked to entanglement entropy or other mixed-state entanglement measures~\cite{christandl_squashed_2004, brandao_faithful_2011, bennett_mixed-state_1996, terhal_entanglement_2002}, but they still characterize the correlation between the bipartitions~\cite{noh_efficient_2020, prosen_operator_2007, prosen_matrix_2009, xu_accessing_2020}.

For the bipartition $[1\dots l]:[(l+1)\dots N]$, there can be up to $d^{2\min\{l, N-l\}}$ non-zero singular values, which still necessitates keeping track of an exponentially large number of parameters to describe an arbitrary state. To avoid the exponential complexity, we omit terms with small singular values and retain only the $\chi$ largest singular values for each cut:
\begin{equation}\label{eq:MPO_truc}
    \rho_{I_1, \dots, I_N} = \sum_{\alpha_1,\alpha_2,\alpha_3, \dots, \alpha_{N-1}=0}^{\chi-1}
    \Gamma^{[1]I_{1}}_{\alpha_{1}} \lambda^{[1]}_{\alpha_{1}} \Gamma^{[2]I_{2}}_{\alpha_{1},\alpha_{2}} \lambda^{[2]}_{\alpha_{2}} \Gamma^{[3]I_{3}}_{\alpha_{2},\alpha_{3}} \dots \lambda^{[N-1]}_{\alpha_{N-1}} \Gamma^{[N]I_{N}}_{\alpha_{N-1}}.
\end{equation}
Here, $\chi$ is called the bond dimension. After truncation, only $\mathcal{O}(N\chi^2)$ parameters are needed to describe the state. Qualitatively speaking, a larger bond dimension allows for the accurate representation of more correlated states, and MPO with $\chi \ge d^{2\lfloor N/2\rfloor}$ is guaranteed for the exact simulation of an arbitrary state. Using the method provided in Ref.~\cite{noh_efficient_2020}, we are able to simulate the Haar random circuit with the depolarizing channel with time complexity of $\mathcal{O}(N^2 \chi^3)$ per timestep of the simulation.

In the previous sections and the main text, we calculated the von Neumann entropies of the subsystems to obtain CMI. However, while determining the entanglement spectra (eigenvalues of the reduced density matrices) is straightforward for stabilizer states and MPS, calculating the von Neumann entropy in MPO simulation is significantly more complex -- even determining whether a given MPO is positive definite is known to be NP-hard~\cite{kliesch_matrix_2014}. To circumvent this challenge, we calculate second-order Rényi entropy,
\begin{equation}\label{eq:renyi}
    S^{(2)}(X) = -\log_{d}\Tr(\rho_{X}^2),
\end{equation}
instead of von Neumann entropy. Here, $\rho_X$ is the reduced density matrix of the subsystem $X$. For a given MPO, calculating Eq.~\ref{eq:renyi} can be done through simple tensor contractions. We then compute CMI in terms of the Rényi entropies,
\begin{equation}
    I^{(2)}(A:C|B) = S^{(2)}(AB) + S^{(2)}(BC) - S^{(2)}(B) - S^{(2)}(ABC).
\end{equation}
Here, the superscript $^{(2)}$ indicates the use of second-order Rényi entropy. It is important to note that stabilizer states have uniform entanglement spectra on their support, so the Rényi entropies of any order, including the von Neumann entropy, have the same value. Thus, this change in the definition of CMI does not affect the results presented in the main text and previous sections of SM.

\subsection{Results of MPO simulation}

\begin{figure}
\includegraphics[width=\textwidth]{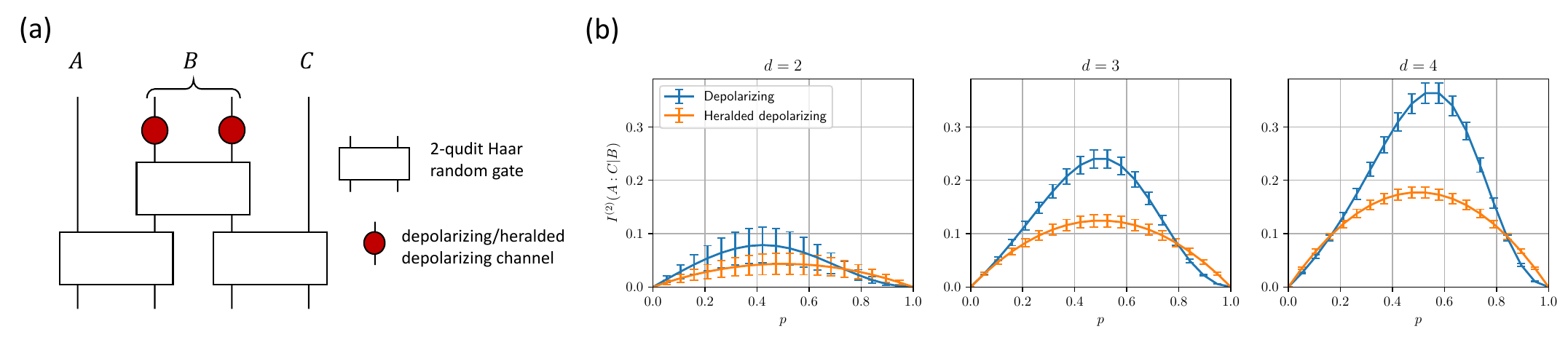}
\caption{\label{fig:toyexample+model}(a) The system starts with the state \(\ket{0}^{\otimes 4}\), and two layers of Haar random gates are applied to adjacent qudits. Depolarizing or heralded depolarizing channels are then applied to the two middle qudits. The first qudit, the two middle qudits, and the last qudit are labeled as $A, B,$ and $C$, respectively, and we compute $I^{(2)}(A:C|B)$. (b) Numerical results of the toy example of the various local dimensions ($d=2,3,4$) are shown for both circuits with depolarizing and heralded depolarizing channels.}
\centering
\end{figure}

\begin{figure}
\includegraphics[width=\textwidth]{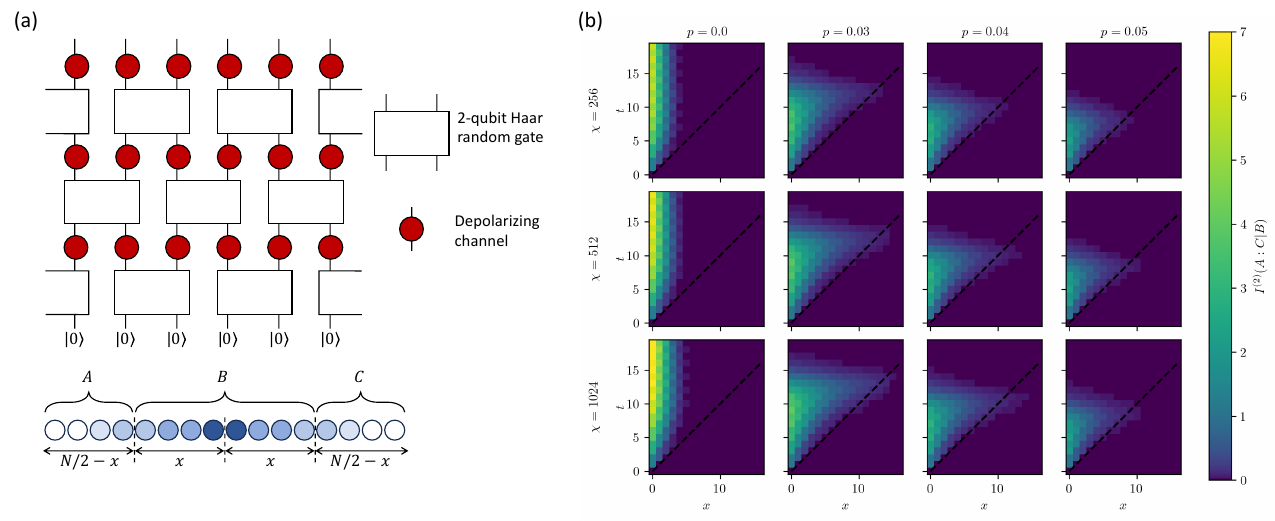}
\caption{\label{fig:MPO_setup+cmap}(a) Haar random circuit with depolarizing channel. The circuit consists of alternating layers of Haar random circuits acting on neighboring qubits, with depolarizing channels applied to each qubit at every timestep with an error rate $p$. At each timestep, the system is divided into $A, B$, and $C$, where $A$ is the first $N/2 - x$ qubits, $B$ is the $2x$ qubits in the middle, and $C$ is the last $N/2 - x$ qubits, and then we calculate $I^{(2)}(A:C|B)$. (b) Results of the spreading of CMI from the MPO simulations. We run the circuits with $32$ qubits and various error rates and bond dimensions, and $I^{(2)}(A:C|B)$ is averaged over $32$ circuit realizations.}
\centering
\end{figure}

\begin{figure}
\includegraphics[width=\textwidth]{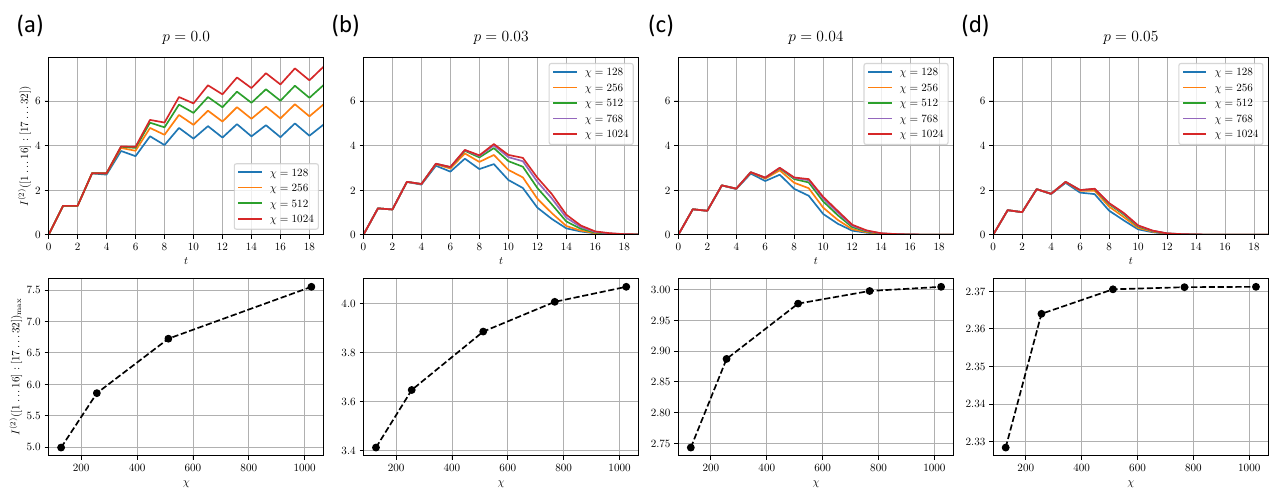}
\caption{\label{fig:MPO_benchmark+label}
Mutual information of equal bipartition $I^{(2)}([1,\dots,16]:[17,\dots,32])$ for benchmarking the MPO simulations with the error rates (a) $p=0$, (b) $p=0.03$, (c) $p=0.04$, and (d) $p=0.05$. Upper panels are the evolution of the mutual information, and the lower panels are the maximum mutual information $I^{(2)}([1,\dots,16]:[17,\dots,32])_{\mathrm{max}}$ achieved in the first $20$ timesteps.}
\centering
\end{figure}

Finally, we provide numerical evidence that the key findings of the main text can be generalized to the Haar random circuits with depolarizing channels. First, to explore whether CMI spreading beyond the lightcone generalizes to Haar random circuits with depolarizing channels, we examine a simple toy example with 4 qudits. As depicted in Fig.~\ref{fig:toyexample+model}(a), we start with 4 qudits prepared in a product state \(\ket{0}^{\otimes 4}\). We then apply 2-qudit Haar random gates to the first and last pairs of qudits, followed by another Haar random gate on the two qudits in the middle. Afterward, we apply either depolarizing channels or heralded depolarizing channels to the two middle qudits. We label the first qudit as $A$, the second and third qudits as $B$, and the last qudit as $C$, and then calculate the CMI, $I^{(2)}(A:C|B)$, averaged over $16$ circuit realizations. Importantly, note that $A$ and $C$ are separated by the lightcone.

The numerical results obtained using MPO for this toy example are displayed in Fig.~\ref{fig:toyexample+model}(b). We compare the CMI for both depolarizing and heralded depolarizing channels across different local dimensionalities. We set the bond dimension $\chi=d^4$, ensuring that the simulations yield exact results. The results confirm that heralding the locations of errors can either increase or decrease CMI, as discussed earlier. Despite these variations, this toy example demonstrates that CMI spreading beyond the lightcone is not unique to the heralded depolarizing channel but is a universal phenomenon applicable to more general noise models, including the depolarizing channel.

Next, we present the numerical results for Haar random circuits with the depolarizing channels. As seen in Fig.~\ref{fig:MPO_setup+cmap}(a), we start with the product state $\ket{0}^{\otimes N}$ and alternately apply 2-qubit Haar random gates on the even and odd pairs of the adjacent qubits. After each layer of the Haar random gates, we apply the depolarizing channel with the error rate $p$ on each qubit. At each timestep, we calculate $I^{(2)}(A:C|B)$, where $A$ is the first $N/2 - x$ qubits, $B$ is the $2x$ qubits in the middle, and $C$ is the last $N/2 - x$ qubits. All numerical results for Haar random circuits are from the MPO simulations with $N=32$ while varying error rates $p$ and bond dimensions $\chi$, with CMIs averaged over $32$ circuit realizations.

Fig.~\ref{fig:MPO_setup+cmap}(b) shows the numerical results for the evolution of CMI under different error rates and bond dimensions. When the circuit is noiseless ($p=0$), we observe that CMI initially spreads linearly but becomes distorted around $t \approx 7$. Ideally, CMI would continue spreading until reaching the system boundary, indicating the limitations of the MPO method for simulating quantum systems with high correlations. Specifically, for noiseless random circuits, correlations grow linearly until they saturate at a value proportional to the system size. Meanwhile, as noted in Sec~\ref{subsec:mpo_intro}, MPOs only represent states with limited correlations, so the MPO method fails for deep, noiseless random circuits. However, introducing noise makes simulation feasible because noise limits the accumulation of correlation. In particular, the maximum correlation in noisy random circuits with a fixed error rate is known to be bounded by a constant, regardless of system size~\cite{noh_efficient_2020, li_entanglement_2023}. Thus, when the bond dimension is sufficiently large for a given error rate, MPO simulations can accurately describe the dynamics.

To benchmark the numerical results more quantitatively, we analyze the amount of correlation in the MPO simulations. Since equal bipartitions are expected to have maximum correlation, we calculate the mutual information between $[1,2,\dots,16]$ and $[17,18,\dots,32]$. Again, we write the mutual information in terms of Rényi entropies,
\begin{equation}
    I^{(2)}([1,\dots,16]:[17,\dots,32]) = S^{(2)}([1,\dots,16]) + S^{(2)}([17,\dots,32]) - S^{(2)}([1,\dots,32]).
\end{equation}
Note that observing the convergence of $I^{(2)}([1,\dots,16]:[17,\dots,32])$ as the bond dimension increases indicates how large the bond dimension should be for accurate simulation. Fig~\ref{fig:MPO_benchmark+label} shows how the mutual information of equal bipartitions evolves and the maximum value of the mutual information achieved in the first 20 timesteps. For the noiseless circuits, $\chi = 1024$ is sufficiently large for an accurate simulation until $t\approx 6$, as $I^{(2)}([1,\dots,16]:[17,\dots,32])$ converges. After $t=7$, however, $\chi=1024$ is not enough for $I^{(2)}([1,\dots,16]:[17,\dots,32])$ to converge, indicating the failure of accurate simulation. On the other hand, for noisy cases, $\chi=1024$ is enough for the accurate simulation given sufficiently large error rates. In particular, for $p=0.04$ and $0.05$, both the evolution of $I^{(2)}([1,\dots,16]:[17,\dots,32])$ and the maximum correlation achieved indicate that the bipartite correlation converges for $\chi=1024$, suggesting accurate simulation results.

Lastly, we analyze the spreading of CMI in noisy circuits presented in Fig.~\ref{fig:MPO_setup+cmap}(b). As noted above, the results from the noiseless circuit until $t\approx 5$ and the noisy circuits with $p\ge 0.04$ are accurately simulated for $\chi=1024$. Based on these credible numerical results, we observe that in the noisy circuit, unlike the noiseless circuits, CMI spreads superlinearly and propagates beyond the lightcone. In conclusion, Haar random circuits with depolarizing channels reproduce the main features observed in Clifford random circuits with heralded depolarizing channels. This suggests that the findings in the main text are not limited to the specific models studied but reflect more general characteristics of CMI in noisy random circuits. Therefore, we expect that the critical behavior and scaling laws observed in the main text are generic and applicable to Haar random circuits with depolarizing channels. However, due to the limitations of small system size and large error rates, we do not numerically observe the critical behavior and scaling laws in the presented results.


\end{document}